\documentclass[useAMS,usenatbib,usegraphicx]{mn2e}
\usepackage{url}

\newcommand{\Ni}{$^{56}$Ni}
\newcommand{\Co}{$^{56}$Co}
\newcommand{\Msun}{$M_\odot$}

\title[SLSN 2006gy]
{
Light~Curve~Modeling~of~Superluminous~Supernova~2006gy:
Collision between Supernova Ejecta and Dense Circumstellar Medium
}
\author[T. J. Moriya et al.]
{Takashi J. Moriya$^{1,2,3}$\thanks{takashi.moriya@ipmu.jp},
 Sergei I. Blinnikov$^{4,5,1}$,
 Nozomu Tominaga$^{6,1}$,
 Naoki Yoshida$^{7,1}$, \newauthor
 Masaomi Tanaka$^{8}$,
 Keiichi Maeda$^{1}$, \&
 Ken'ichi Nomoto$^{1}$ \\
$^{1}$
Kavli Institute for the Physics and Mathematics of the Universe,
Todai Institutes for Advanced Study,
University of Tokyo, \\ 5-1-5 Kashiwanoha, Kashiwa, Chiba 277-8583, Japan
\\
$^{2}$
Department of Astronomy, Graduate School of Science, University of
Tokyo, 7-3-1 Hongo, Bunkyo, Tokyo 113-0033, Japan \\
$^{3}$
Research Center for the Early Universe, Graduate School of Science, University of Tokyo,
7-3-1 Hongo, Bunkyo, Tokyo 113-0033, Japan\\
$^{4}$
Institute for Theoretical and Experimental Physics, Bolshaya Cheremushkinskaya 25, 117218 Moscow, Russia\\ 
$^{5}$
Sternberg Astronomical Institute, Moscow University, Universitetski pr. 13, 119992 Moscow, Russia\\
$^{6}$
Department of Physics, Faculty of Science and Engineering, Konan University, 8-9-1 Okamoto, Kobe, Hyogo 658-8501, Japan\\
$^{7}$
Department of Physics, Graduate School of Science, University of
Tokyo, 7-3-1 Hongo, Bunkyo, Tokyo 113-0033, Japan \\
$^{8}$
National Astronomical Observatory of Japan, 2-21-1 Ohsawa, Mitaka, Tokyo 181-8588, Japan\\
}
\begin{document}

%\date{Accepted 1988 December 15. Received 1988 December 14; in original form 1988 October 11}

\pagerange{\pageref{firstpage}--\pageref{lastpage}} \pubyear{2012}

\maketitle

\label{firstpage}

\begin{abstract}
We show model light curves of superluminous supernova 2006gy
on the assumption that the supernova is powered by the collision 
of supernova ejecta and its dense circumstellar medium.
The initial conditions are constructed based on the shock 
breakout condition, assuming that the circumstellar medium is dense 
enough to cause the shock breakout within it. 
We perform a set of numerical light curve calculations by using 
a one-dimensional multigroup radiation hydrodynamics 
code \verb|STELLA|.
We succeeded in reproducing the overall features 
of the early light curve of SN 2006gy with the circumstellar medium
whose mass is about $15~M_\odot$
(the average mass-loss rate $\sim 0.1~M_\odot~\mathrm{yr^{-1}}$).
Thus, the progenitor of SN 2006gy is likely a very massive star.
The density profile of the circumstellar medium is not well
constrained by the light curve modeling only, but our modeling 
disfavors the circumstellar medium formed by steady mass loss.
The ejecta mass is estimated to be comparable
to or less than $15~M_\odot$ and the explosion energy is expected 
to be more than $4\times 10^{51}$ erg.
No \Ni\ is required to explain the early light curve.
We find that the multidimensional effect, e.g.,
the Rayleigh-Taylor instability, which is expected
to take place in the cool dense shell between the supernova ejecta
and the dense circumstellar medium, is important in understanding 
supernovae powered by the shock interaction. 
%Further investigations of the collision
%in multidimension are required for the better understanding of
%superluminous supernovae.
We also show the evolution of the optical and near-infrared
model light curves of high-redshift superluminous supernovae.
They can be potentially used to identify SN 2006gy-like superluminous 
supernovae in the future optical and near-infrared transient surveys.
\end{abstract}


%%%%%%%%%%%%%%%%%%%%
\section{Introduction}\label{sec1}
Recent untargeted transient surveys such as
Texas Supernova Search \citep{quimby2006b},
Palomar Transient Factory \citep{law2009,rau2009},
Catarina Real-time Transient Survey \citep{drake2009},
Panoramic Survey Telescope \& Rapid Response System
\citep{hodapp2004},
discovered new kinds of supernovae (SNe).
Among the most specutaclar are superluminous SNe (SLSNe)
which become typically brighter than $\sim -21$ mag in optical
\citep[e.g.,][]{smith2007b,quimby2007,quimby2011,rest2011,miller2009,gezari2009,
gal-yam2009,drake2010,drake2011,chatzopoulos2011b,pastorello2010,barbary2009,chomiuk2011,leloudas2012,berger2012}.

SN 2006gy is the first example of SLSNe with detailed
photomentric and spectroscopic observations.
The luminosity of SN 2006gy stayed brighter than $-21$ mag in the $R$ band
for more than 50 days (see Section \ref{observations} for a brief summary of observations).
It was the most luminous SN ever reported at that time.
Many theoretical models have been proposed to explain the huge luminosity of SN 2006gy.
SN 2006gy was, at first, linked to Type Ia SNe
exploding inside H-rich circumstellar medium (CSM)
\citep[e.g., SN 2002ic;][]{hamuy2003,deng2004} \citep{ofek2007}.
However, the total observed radiation energy exceeded the available
energy from a Type Ia explosion and thus the notion was adandoned.

Currently, SN 2006gy is thought to be caused by the death of 
a massive star. Several models have been proposed to explain the
observed features as follows. All of them posit a massive 
progenitor star.

\begin{enumerate}
{\bf \item Large production of \Ni}

The energy released by the radioactive decay of \Ni\ is a common 
luminosity source of SNe.
However, more than $10~M_\odot$ of \Ni\ would be required
to account for the peak luminosity of SN 2006gy \citep[e.g.,][]{nomoto2007}.
The large amount of \Ni\ is not easily produced even by
very high energetic core-collapse SNe \citep{umeda2008}.
Pair-instability SNe \citep[PISNe, e.g.,][]{barkat1967,rakavy1967}
can produce the large \Ni\ but
the rising time of SN 2006gy is found to be much shorter
than that expected from a PISN
\citep[e.g.,][]{kasen2011,nomoto2007,moriya2010}.
The weakness of Fe lines in spectra at the late phase
is also against the \Ni\ heating scenario \citep{kawabata2009}.

{\bf \item Interaction between SN ejecta and dense CSM}

The interaction between SN ejecta and
dense CSM created by the progenitor before the explosion
can make the luminosity of an SN very large.
Strong shocks convert the kinetic energy of SN ejecta 
to thermal energy which is eventually
released as radiation energy.
\citet{smith2007} suggest that the declining phase of the light curve (LC) of SN 2006gy
can be explained by emission from a shocked $\sim 10~M_\odot$ CSM shell.
\citet{agnoletto2009} combined the \Ni\ and shock
interaction to explain the early LC of SN 2006gy.
They argue that the rising part of LC is powered by \Ni\ decay and
the shock interaction comes into play from around the LC peak.
Then, the required \Ni\ mass is reduced to a few \Msun.

{\bf \item Pulsational pair-instability}

\citet{woosley2007} relate SN 2006gy to the pulsational pair-instability. 
A very massive star with mass $\simeq 80-130~M_\odot$ 
at the zero-age main sequence % for zero metallicity
undergoes strong pulsations \citep[see also][]{umeda2008,ohkubo2009},
which induce the intermittent extensive mass loss.
If the mass ejected by such an eruption 
is caught up by the mass released by the next eruption,
the two massive shells collide and the star can eventually be as luminous
as SN 2006gy.
In this case, the luminosity source is the kinetic energy of the 
secondly ejected materials. Hence, the radiation mechanism is 
essentially the same as that of (ii) the interaction between
SN ejecta and dense CSM. The ejecta from inside is released by 
the pulsational pair-instability, rather than an SN explosion in this case.
This kind of luminous transients are called pulsational
pair-instability SNe.

{\bf \item 
Spin-down of newly born magnetars}

If a magnetar born at the time of the core collapse of a massive star
has suitable magnetic field and spin, it can release its rotational
energy efficiently just after the SN explosion.
If the radiation energy released at the magnetar can be converted to thermal
energy, SN ejecta surrounding the magnetar is heated up and the SN can
become very bright (\citealt{kasen2010,woosley2010}, see
also \citealt{maeda2007}).

{\bf \item
Transition of a neutron star to a quark star}

Neutron in a neutron star formed at the core collapse can be
further discomposed to quarks and they can form a quark star.
When a neutron transforms to quarks,
latent heat due to the phase transition can heat up the SN ejecta
surrounding the quark star and make an SLSN
(e.g., \citealt[][]{leahy2008,ouyed2010,kostka2012}, see also \citealt{benvenuto1999}).

\end{enumerate}

Among the possible heating sources, the CSM interaction
(ii and iii) is plausible because the spectra of SN 2006gy show 
narrow P-Cygni profiles
that indicate the existence of a CSM outflow with $\sim 100~\mathrm{km~s^{-1}}$.
Although the observed low X-ray flux of SN 2006gy 
(see Section \ref{observations})
might appear to contradict the interaction scenario,
the CSM could be optically thick enough to
absorb X-rays from the shock wave, or
the shock wave itself could be cool
\citep[e.g.,][]{blinnikov2008,chevalier2012,svirski2012}.

Although the scenario (ii), i.e.,
the shock interaction between SN ejecta and dense CSM, is
suggested to be able to explain the early LC of SN 2006gy,
detailed LC models have not been explored yet.
\citet{chevalier2011,moriya2011,svirski2012} investigate the possibility of
explaining SN 2006gy by the interaction of SN ejecta and dense CSM alone.
They assume that the CSM is optically thick enough to cause
the shock breakout within it
(e.g., \citealt[][]{weaver1976,nakar2010}, see also \citealt{ofek2010,balberg2011}
about the shock breakout in dense CSM).
\citet{chatzopoulos2011} show the bolometric LC from the interaction scenario
based on their semi-analytic LC model (see Section \ref{semi}).
The previous works regarding the collision of SN ejecta and dense CSM
so far analyse the LC of SN 2006gy
based on analytic approaches with several simplifications.
The detailed numerical modeling of the LC of SN 2006gy has not been done
yet and it is clearly required
(see also the very recent attempt by \citealt{ginzburg2012}).

In this paper, we study model LCs for SN 2006gy by using
numerical radiation hydrodynamics code \verb|STELLA|. 
Our purpose of the modeling is to show numerically that the simple
analytic shock breakout model actually works well
to explain overall features of the early LC of SLSN 2006gy.
We focus on SN 2006gy because it is the best observed
SLSNe currently reported.
We do not aim at the perfect fitting of the LC
because it does not necessarily lead us to the correct physical
parameters of SN 2006gy.
This is because
\verb|STELLA| makes several simplifications to numerically treat
the radiation hydrodynamics.
We rather concentrate on the overall features of the LC of SN 2006gy
and show that the dense CSM configurations predicted by
the shock breakout model are actually working
well to reconstruct the LC of SN 2006gy.

The rest of the paper is organized as follows.
Section \ref{observations} is a short summary of the observations of SN 2006gy.
In Section \ref{method}, \verb|STELLA|, which is used for our LC modeling,
is summarized briefly. How multidimensional effect is included in
one-dimensional code \verb|STELLA| is also overviewed in Section \ref{method}.
Our models are introduced in Section \ref{sec2}
and the LC calculations based on the models are shown in Section \ref{LC}.
Discussion is given in Section \ref{discussion}.
We close the paper with conclusions in Section \ref{conclusions}.
We use the standard cosmology with
$H_0=70~\mathrm{km~s^{-1}~Mpc^{-1}}$,
$\Omega_{M}=0.3$, and
$\Omega_{\Lambda}=0.7$ when it is required.

\section{Brief Summary of Observations}\label{observations}
We briefly summarize the observational properties of SN 2006gy.
SN 2006gy was discovered by Texas Supernova Search
on September 18, 2006 (UT) near the nucleus of
an early-type galaxy NGC 1260 \citep{quimby2006}.
There are several suggested values for 
the extinction by the host galaxy
\citep{ofek2007,smith2007b,agnoletto2009}.
We adopt $E(B-V)_\mathrm{host}=0.40$ mag following
\citet{agnoletto2009} with the Milky Way extinction
$E(B-V)_\mathrm{MW}=0.16$ mag \citep{schlegel1998}.
Thus, the total extinction is $E(B-V)=0.56$ mag
or $A_R=1.3$ mag with $R_V=3.1$ \citep{cardelli1989}.
The distance modulus $\mu$ of the host galaxy is 
also taken from \citet{agnoletto2009} ($\mu=34.53$ mag).

Follow up spectral observations classified SN 2006gy
as Type IIn because of the narrow H emission lines presented in
the spectra \citep[see][for the details of Type IIn]{schlegel1990,filippenko1997}.
The luminosity of SN 2006gy kept rising until October 25, 2006
(UT) and the peak $R$ band luminosity got close to $\simeq -22$ mag.
The rising time is estimated as about 70 days
(hereafter, days are in the rest frame).
After reaching the peak luminosity, the LC declined slowly
($\simeq 0.02~\mathrm{mag~day^{-1}}$)
for $\simeq 120$ days and then the LC stayed almost constant
for $\simeq 20$ days until SN 2006gy hid behind the Sun.
Near-infrared (NIR) LCs in these early epochs are consistent with
the blackbody temperature obtained from the optical spectra
and no significant excess was detected \citep{miller2010}.
No X-ray was detected when the LC is rising \citep{ofek2007}
but weak X-rays may have been detected during the declining phase \citep{smith2007b}.
No radio emission is detected at any observed epochs
\citep{ofek2007,chandra2007,argo2007,bietenholz2007,bietenholz2008a,bietenholz2008b}.

Optical spectra of the early epochs are also taken intensively
\citep[e.g.,][]{smith2010,agnoletto2009}.
Spectra taken before the LC peak are characterized by
Lorentzian H$\alpha$ emission lines \citep{smith2010}.
The origin of the Lorentzian profile is related to
the existence of optically thick
CSM \citep[e.g.,][]{chugai2001,dessart2009}.
Except for the narrow H emission lines, the spectra are featureless
and characterized by blackbody with depletion in blue \citep{smith2010,agnoletto2009}.
The reason for the
lack of features may be partly because the spectra before the LC peak
were taken only with low resolutions.
After the LC peak, overall H$\alpha$ line profiles can be fitted by
two Gaussian components with the FWHM of 1,800 $\mathrm{km~s^{-1}}$
and 5,200 $\mathrm{km~s^{-1}}$ and they are presumed to come from
the interacting region between the ejecta and dense CSM
\citep[e.g.,][]{smith2010}.
There also exists a broad absorption in the blue part of the
H$\alpha$ profile which is suggested to originate from the ejecta inside.
In addition, the spectra of SN 2006gy show narrow P-Cygni profiles from
several elements (e.g., H, Fe)
with the outflowing velocity $\sim 100~\mathrm{km~s^{-1}}$.
As the velocity is too slow to attribute it to the ejecta inside,
those narrow lines are presumed to originate from the unshocked CSM.
The strengths of these narrow lines decline with time
and they are barely seen in the spectrum taken
at $\simeq$ 140 days since the LC peak \citep{smith2010}.

About 100 days later, SN 2006gy came out of the Sun and
was observed again \citep{agnoletto2009,kawabata2009,miller2010}.
The optical luminosity of SN 2006gy declined dramatically (about 2 mag
in the $R$ band) which was almost constant for $\simeq 20$ days before the SN
went behind the Sun. 
The luminosity declined very slowly $(\simeq 0.002~\mathrm{mag~day^{-1}})$
since it appeared from the Sun
for more than 400 days until the last reported observation on
November 22, 2008 (UT)
\citep{kawabata2009,miller2010}.
The decline rate 
is much slower than that of \Co\ decay $(0.01~\mathrm{mag~day^{-1}})$
and the main source of the luminosity cannot be the
\Co\ decay. Because of the high NIR luminosities, 
\citet{smith2008,miller2010} suggest that 
the late time luminosity is due to light echoes.
The optical spectra of those epochs are dominated by
intermediate width emission lines 
\citep[$\simeq2,000~\mathrm{km~s^{-1}}$,][]{kawabata2009}.
H emission lines were weaker than those observed in previous epochs
and suggest that the interaction is weak in those epochs and
it is no longer a main source of the radiation
\citep{agnoletto2009,kawabata2009}.
Weakness of Fe lines in those epochs seems 
inconsistent with the large \Ni\ production \citep{kawabata2009}.

\section{Numerical Method}\label{method}
We briefly summarise the basics of \verb|STELLA| code,
which we use for our LC modeling.
We then describe in detail a key numerical method of 
\verb|STELLA| that treats smearing due to
multidimensional effect.

\subsection{STELLA}
\verb|STELLA| is a one-dimensional multigroup radiation
hydrodynamics code
\citep[e.g.,][]{blinnikov1993,blinnikov1998,blinnikov2006,blinnikov2011}
and calculates the spectral energy distributions
(SEDs) at each time step. Multicolor LCs can be obtained by
convolving filter functions to the SEDs.
All the calculations are
performed by adopting 100 frequency bins from $\mathrm{1~\AA}$ to
$\mathrm{5\times10^{4} \AA}$ in log scale.
\verb|STELLA| implicitly treats time-dependent equations
of the angular moments of intensity averaged over a frequency bin.
Local thermodynamic equilibrium is assumed to determine the
ionization levels of materials.
\verb|STELLA| has been intensively used for modeling the SN LCs powered by
the shock interaction
\citep[e.g.,][]{chugai2004,woosley2007,blinnikov2010,moriya2011b}
as well as other types of SNe
\citep[e.g.,][]{baklanov2005,folatelli2006,tsvetkov2009,tominaga2011}.
Comparisons of \verb|STELLA| with other numerical codes are provided in,
e.g., \citet{blinnikov1998,blinnikov2000,blinnikov2003,woosley2007b} and
analytical models are also compared to the numerical results of \verb|STELLA|
\citep[e.g.,][]{rabinak2011}.

\begin{table*}
\centering
\begin{minipage}{160mm}
\caption{List of LC models}
\label{table1}
\begin{tabular}{cccccccccccc}
\hline
Name&$v_s$ & $M_\mathrm{ej}$ & $E_\mathrm{ej}$ & $w$ & $R_o$ & $y_1R_o$ & $xR_o$
& $R_i$ & $M_\mathrm{CSM}$ & $\left<\dot{M}\right>$\footnote{
Average mass-loss rate. $\left<\dot{M}\right>=v_wM_\mathrm{CSM}/(R_o-R_i)$
where $v_w=100~\mathrm{km~s^{-1}}$ is the assumed CSM velocity.}  & $B_q$\\
& $\mathrm{km~s^{-1}}$    &   $M_\odot$     & $10^{51}$ erg & & $10^{15}$ cm & $10^{15}$ cm&
$10^{15}$ cm&$10^{15}$ cm& $M_\odot$ & $M_\odot~\mathrm{yr^{-1}}$ & \\
\hline
A1 &5,200& 20 & 50 & 5 & 11  & 4.9 & 1.8 & 1.0 & 22 & 0.70 & 1 \\
A2 &5,200& 20 & 50 & 5 & 11  & 4.9 & 1.8 & 1.5 & 10 & 0.33 & 1 \\
B1 &5,200& 20 & 10 & 2 & 8.8 & 3.3 & 0.090& 0.090& 0.83 & 0.030 & 1\\
B2 &5,200& 20 & 30 & 2 & 8.8 & 8.3 & 2.0 & 0.090 & 27 & 0.98 & 1 \\
C1 &5,200& 20 & 50 & 0 & 4.9 & 4.8 & 1.8 & 1.0 & 14 & 1.1 & 1 \\
\hline
D1 &10,000& 20 & 10 & 5 & 21 & 11 & 4.6 & 4.5 & 22 & 0.42 & 1\\
D2 &10,000& 20 & 10 & 5 & 21 & 11 & 4.6 & 5.0 & 18 & 0.36 & 1\\
E1 &10,000& 20 & 30 & 2 & 17 & 6.3& 0.32 & 0.10 & 3.2 & 0.060 & 1 \\
E2 &10,000& 2 & 10 & 2 & 17 & 6.3& 0.32 & 0.10 & 3.2 & 0.060 & 1 \\
F1 &10,000& 20 & 10 & 0 & 11 & 11 & 4.6 & 5.0 & 15 & 0.45 & 1 \\
D3 &10,000& 10 & 10 & 5 & 21 & 11 & 4.6 & 5.0 & 18 & 0.36 & 1\\
D4 &10,000& 30 & 10 & 5 & 21 & 11 & 4.6 & 5.0 & 18 & 0.36 & 1\\
D5 &10,000& 10 & 5  & 5 & 21 & 11 & 4.6 & 5.0 & 18 & 0.36 & 1\\
D6 &10,000& 20 & 10 & 5 & 21 & 11 & 4.6 & 5.0 & 18 & 0.36 & 0.33\\
D7 &10,000& 20 & 10 & 5 & 21 & 11 & 4.6 & 5.0 & 18 & 0.36 & 3\\
\hline
\end{tabular}
\end{minipage}
\end{table*}

\subsection{Smearing}\label{sec2:sm}
As is shown in the previous works \citep[e.g.,][]{chevalier1994,chugai2001,chugai2004}
and will be shown in the following sections,
the interaction of SN ejecta and dense CSM
results in a dense cool shell between SN ejecta and dense CSM.
This is because of the radiative cooling of the shocked region.
The shocked region becomes very dense and cools down efficiently by radiation.
The cooling prevents the pressure
from growing sufficiently enough to sustain the shell.
Hence, the shell becomes thinner and denser and the cooling
becomes more efficient. Thus, this cooling process
is catastrophic.
However, in reality, 
such a shell is unstable because of several instabilities
like the Rayleigh-Taylor instability which require multidimensional
calculations to treat \citep[e.g.,][]{chevalier1995}.
The multidimensional effect smears the shell and less kinetic energy is
converted to radiation energy.
In other words, the cooling by radiation is less efficient in three dimensions
than in one dimension.
In \verb|STELLA| code, we take such multidimensional effects into account
by introducing a smearing term in the equation of motion
so that the conversion efficiency from kinetic energy to radiation
energy
can be reduced \citep{blinnikov1998}.
This term is similar to artificial viscosity,
although the smearing term has completely opposite effect.

As is shown in \citet{blinnikov1998}, the smearing term
is defined such that the total energy is manifestly conserved.
Only the neighbouring zones are affected by the smearing term.
The overall normalization factor $R_\mathrm{cut}(\tau)$ of the smearing
is expressed as 
\begin{equation}
R_\mathrm{cut}(\tau)=B_qf(\tau).
\end{equation}
See \citet{blinnikov1998} for
the definitions of $R_\mathrm{cut}(\tau)$ and the smearing term.
$f(\tau)$ is introduced so that the artificial smearing 
is reduced at optically thick regions where the effect of cooling
is less efficient and the multidimensional instabilities due to the
cooling grow less.
In \verb|STELLA| code, $f(\tau)$ is
an empirically obtained monotonically decreasing function
which satisfies $f(\tau\rightarrow 0)=1$.
$B_q$ determines the overall strength of the smearing effect.
Ideally, $B_q$ should be calibrated by comparing results obtained by
our one-dimensional calculations to those of multidimensional
calculations in which the effect of multidimensional instabilities
in the shell are taken into account.
However, such a multidimensional radiation
hydrodynamical code with which we can compare our results
is not available yet.
We use $B_q=1$ as our standard value.
We also show the effect of $B_q$ on model LCs (Section \ref{sm} and
Appendix \ref{appendix}).
The results of LC calculations strongly depend on $B_q$,
as the smearing term directly affects the conversion efficiency
from kinetic energy to radiation.
We need a multidimensional radiation hydrodynamics code
for the calibration of the parameter $B_q$.

\section{Initial Conditions}\label{sec2}
In this section, we show how the initial conditions
for our LC calculations are constructed.
Two components exist in the initial conditions:
SN ejecta inside and CSM outside.
We assume that there is a gap between the progenitor and the dense CSM.
Then, it takes some time for the SN ejecta to reach the dense CSM and
start to collide.
We assume that the SN ejecta freely expands in the gap before the collision.
We numerically follow the LCs after the collision.
The initial conditions of the two components, SN ejecta and dense CSM,
are constructed by the way explained in this section.
Then, these initial conditions will be confirmed
by our numerical LC calculations as we show in the later sections.
The initial density structures of two representative
models are shown in Figure \ref{density} as examples.
Both SN ejecta and CSM are assumed to have solar metallicity
and no $^{56}\mathrm{Ni}$ is included in our calculations
unless it is otherwise mentioned.
The summary of the models is given in Table \ref{table1}.

%To constrain possible ranges of input parameters in our calculations,
%we first construct the density structures of these two components
%from observational properties by analytically interpreting them.
%Then, these initial conditions will be confirmed
%by our numerical LC calculations as we show later.

\subsection{Supernova Ejecta}\label{iniSN}
SN ejecta before the collision is assumed to be freely expanding
with a homologous velocity profile.
The analytic approximation for the density structure of SN ejecta
provided by, e.g., \citet{chevalier1989} is adopted:
\begin{equation}
{\Large
\rho\left(r,t\right)=\left\{ \begin{array}{ll}
\zeta_\rho\frac{M_\mathrm{ej}}{v_\mathrm{t}^3t^3}\left(\frac{r}{v_\mathrm{t}t}\right)^{-\delta} &
 (v<v_\mathrm{t}), \\ \\
\zeta_\rho\frac{M_\mathrm{ej}}{v_\mathrm{t}^3t^3}\left(\frac{r}{v_\mathrm{t}t}\right)^{-n} & (v>v_\mathrm{t}), \\
\end{array} \right.}
\end{equation}
where $M_\mathrm{ej}$ is the SN ejecta mass, $t$ is time since the explosion,
$v_\mathrm{t}=\zeta_v\left(E_\mathrm{ej}/M_\mathrm{ej}\right)^{1/2}$, and
$E_\mathrm{ej}$ is the kinetic energy of the SN ejecta.
The constants $\zeta_\rho$ and $\zeta_v$ are constrained by the condition
that the sum of density and kinetic energy should be $M_\mathrm{ej}$ and $E_\mathrm{ej}$:
\begin{eqnarray}
 \zeta_v&=&\left[\frac{2\left(5-\delta\right)\left(n-5\right)}
{\left(3-\delta\right)\left(n-3\right)}\right]^{1/2}, \\
 \zeta_\rho&=&\frac{1}{4\pi}
\frac{\left(n-3\right)\left(3-\delta\right)}{n-\delta}.
\end{eqnarray}
In this paper, we adopt $\delta=1$ and $n=7$, which is used by
\cite{chevalier2011}.
We have also tried $n=6$ and $n=8$ for some models but results
had little difference compared to those of $n=7$.
The maximum velocity of the SN ejecta before the interaction
is chosen to be high enough,
so that most of the assumed $E_\mathrm{ej}$ is contained in the SN ejecta.
It is around $20,000 - 50,000~\mathrm{km~s^{-1}}$.

$M_\mathrm{ej}$ is difficult to be constrained only by
the observations of the LC of SN 2006gy
because the LC is mainly affected by CSM as is shown
in the following sections.
In most of the models, we adopt $M_\mathrm{ej}=20~M_\odot$
because the progenitors of Type IIn SNe are presumed to be originated
from relatively massive stars.
Effect of $M_\mathrm{ej}$ on LCs is discussed
in Sections \ref{sec:mejeej} and \ref{sec:eff}.

\subsection{Circumstellar Medium}\label{iniCSM}
Dense CSM in the calculations is assumed to exist from
$R_i$ to $R_o$ from the center and
have a density structure $\rho\propto r^{-w}$.
The outflowing velocity of the CSM is $100~\mathrm{km~s^{-1}}$.
It is estimated from the narrow P-Cygni profile of H$\alpha$
appeared in the spectra of SN 2006gy \citep[e.g.,][]{smith2010}.
The CSM is assumed to be optically thick enough to cause the shock breakout
within it.
We estimate the physical conditions of the CSM from the observations by using
the shock breakout condition described in \citet{moriya2011}.
Our main purpose of this paper is to see how well
the properties of the dense CSM predicted by
this simple shock breakout model in CSM
can explain the overall LC features of SN 2006gy.
We use following three values which can be estimated from the observations
to derive the CSM properties:
the photon diffusion time $t_d$ in CSM,
the propagation time $t_s$ of the forward shock through CSM,
and the forward shock velocity $v_s$.
$t_d$ corresponds to the rising time of the LC and
$t_s$ corresponds to 
the time when the narrow P-Cygni H$\alpha$ profiles from CSM disappears.
We adopt $t_d=70$ days and $t_s=194$ days \citep{moriya2011}.
$v_s$ can be estimated from the spectral evolution.

With $t_d$, $t_s$, and $v_s$, we can estimate
the outer radius $R_o$ of the CSM and the radius $xR_o$
where the shock breakout occurs ($R_i/R_o<x<1$)
for a given $w$ based on the shock breakout model.
The shock breakout condition predicts the following relations
for the three values \citep{moriya2011}:
\begin{eqnarray}
t_d&\simeq&
\left\{
\begin{array}{lll}
{\Large
\frac{R_o}{v_s}\left[\left(\frac{c/v_s+ x^{1-w}}{c/v_s+1}\right)^{\frac{1}{1-w}}-x\right]}&& (w\neq1),\\ \\
\frac{R_o}{v_s}\left(x^{\frac{1}{1+c/v_s}}-x\right) && (w=1),
\end{array} \right. \label{td} \\
t_s&\simeq&\frac{R_o-xR_o}{v_s}. \label{ts}
\end{eqnarray}

We try three values for $w$: $w=0,2,5$.
The models with $w=2$ corresponds to the case of the steady mass loss
and they are naturally expected structures for CSM.
Note that X-ray observations of Type IIn SNe suggest that
CSM around Type IIn SN progenitors often does not have
$w=2$ density structures and most of the CSM is likely
from non-steady mass loss \citep[e.g.,][]{dwarkadas2011b}.
A steep CSM density gradient with $w=5$
is suggested for SN 2006gy in \citet{moriya2011}.
We also show $w=0$ models which are difficult to be excluded
only by the LC modeling.

It turns out in the later sections
that it is difficult to estimate $v_s$ from observational values
self-consistently.
This is partly because $v_s$ is not an independent parameter
and, in principle, can be derived for a given CSM structure
if we specify $E_\mathrm{ej}$ and $M_\mathrm{ej}$.
However, $v_s$ is also strongly affected by the conversion
efficiency from the kinetic energy
to radiation through the interaction and it is unknown at first.
Thus, it is difficult to estimate $v_s$ from the first principles.
Hence, we set $v_s$ as a free parameter in this paper.
At first, we try to estimate it from the observations.
As the blackbody radius of SN 2006gy expands linearly with the velocity
$5,200~\mathrm{km~s^{-1}}$, one may estimate that $v_s=5,200~\mathrm{km~s^{-1}}$.
However, the required $E_\mathrm{ej}$ for the
$v_s=5,200~\mathrm{km~s^{-1}}$ models to explain the peak luminosity of
SN 2006gy is found to be very high and it becomes inconsistent
with the relatively low $v_s$.
In other words, the SLSN models obtained by setting
$v_s=5,200~\mathrm{km~s^{-1}}$ are not self-consistent.
Thus, we also try models with higher $v_s$, namely, $v_s=10,000~\mathrm{km~s^{-1}}$.
The $v_s=10,000~\mathrm{km~s^{-1}}$ models are found to work well
self-consistently as is shown in the following sections.
In addition, the linear evolution of the blackbody radius with $5,200~\mathrm{km~s^{-1}}$
is found to be able to be explained by the $v_s=10,000~\mathrm{km~s^{-1}}$ models.
With $t_d=70$ days, $t_s=194$ days, and the given $v_s$,
we can derive $R_o$ and $xR_o$ from Equations (\ref{td}) and (\ref{ts})
for a specified $w$.
In the rest of this section, we show the details of the
two $v_s$ models.

\begin{figure}
\begin{center}
 \includegraphics[width=\columnwidth]{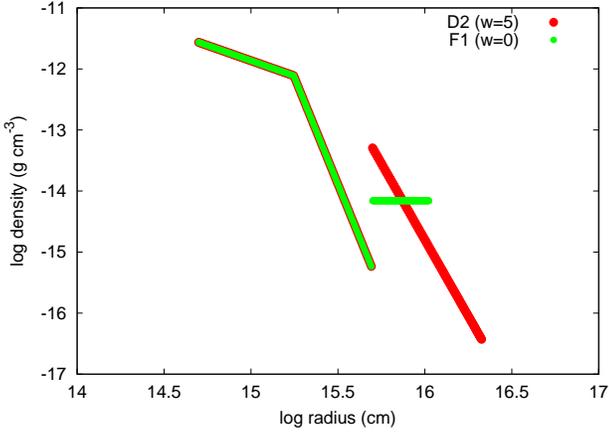}
  \caption{
The density structures of the models D2 ($w=5$) and F1 ($w=0$)
before the collision. These density structures are set as the initial
conditions of the numerical LC calculations.
The structure inside the density jump corresponds to
the SN ejecta which is assumed to be freely expanding
before the collision (Section \ref{iniSN}).
The dense CSM exists above the density jump.
The initial condition of the dense CSM is obtained based on
the shock breakout model within the CSM (Section \ref{iniCSM}).
}
\label{density}
\end{center}
\end{figure}

\subsubsection{$v_s=5,200~\mathrm{km~s^{-1}}$ Model}
This model corresponds to the SN 2006gy model in \citet{moriya2011}.
The shock velocity $v_s=5,200~\mathrm{km~s^{-1}}$ is
estimated from the observed evolution of the blackbody radius of SN 2006gy
\citep{smith2010}.
For $w=5$, we get $R_o=1.1\times10^{16}$ cm and
$xR_o=1.8\times10^{15}$ cm from Equations (\ref{td}) and (\ref{ts}).
The inner radius $R_i$ of CSM cannot be constrained by the above
observables and it is a free parameter.
Two $R_i$ are tried: $10^{15}$ cm (A1) 
and $1.5\times 10^{15}$ cm (A2).
With $w=2$ (steady mass loss), we get
$R_o=8.8\times 10^{15}$ cm and $xR_o=9.0\times 10^{13}$ cm (B1).
We set $R_i=xR_o$ in the $w=2$ models.
The results do not depend so much on $R_i$ in this case
because most of the mass in the $w=2$ CSM
is distributed in the outer part of the CSM.
We also calculate LCs from the model (B2) in which the CSM mass is artificially increased
30 times of the model B1.

In $w=0$ models, $t_s$ becomes similar to $t_d$ $(t_s\simeq t_d)$
\citep{moriya2011}.
Thus, the entire CSM is shocked with $t_d$ and no unshocked wind remains
after the LC peak.
This is against the observations of SN 2006gy because narrow P-Cygni profiles are
observed after the LC peak.
However, this is only true when we only think a single $w=0$ CSM component.
If there is another CSM component outside the main CSM
which is not dense enough to affect the LC but the spectra,
SN 2006gy-like SNe can appear.
Thus, $w=0$ models are difficult to be excluded only by the LC.
From the rising time of the LC, we can presume
 $y_1R_o-xR_o\simeq R_o-xR_o\simeq v_s t_d=3.1\times10^{15}$ cm.
$y_1R_o$ is the radius where the optical depth from the surface of the CSM
becomes $1$ and $y_1R_o\simeq R_o$ when $w=0$ (see \citealt{moriya2011} for the details).
We set the last scattering surface of the $w=5$ model
($y_1R_o=4.9\times 10^{15}$~cm) as $R_o$
so that we can compare the results with those of the $w=5$ models.
Thus, $xR_o=1.8\times 10^{15}$ cm
is also the same as the $w=5$ model and we adopt $R_i=10^{15}$ cm (C1).

\subsubsection{$v_s=10,000~\mathrm{km~s^{-1}}$ Model}
%This model is investigated because it turns out that the models with
%$v_s=5,200~\mathrm{km~s^{-1}}$ require very high $E_\mathrm{ej}$
%(typically $E_\mathrm{ej}\simeq 5\times 10^{52}$ erg, see Section \ref{sec:05})
%to be as luminous as SN 2006gy.
%Assuming relatively low $v_s=5,200~\mathrm{km~s^{-1}}$
%is not compatible with such a high $E_\mathrm{ej}$.
%In addition,
%the velocity obtained from the evolution of the blackbody radius
%from which $v_s=5,200~\mathrm{km~s^{-1}}$ is estimated
%does not necessarily correspond to $v_s$ (see Section \ref{sec:dynamical}).

$v_s=10,000~\mathrm{km~s^{-1}}$ models are constructed by following the
same way as the $v_s=5,200~\mathrm{km~s^{-1}}$ models.
With $t_d=70$ days, $t_s=194$ days, and $v_s=10,000~\mathrm{km~s^{-1}}$,
$R_o=2.1\times10^{16}$~cm and $xR_o=4.6\times 10^{15}$~cm are obtained
for $w=5$. We try two $R_i$: $4.5\times 10^{15}$~cm (D1)
and $5\times10^{15}$~cm (D2 - D7).
Although $R_i=5\times 10^{15}$~cm is slightly larger than the shock breakout
radius $xR_o=4.6\times 10^{15}$~cm,
it turns out that the model gets closer to the SN 2006gy LC.
Given the approximated way of our estimations, 
the difference is within an acceptable range.
The steady mass-loss models
($w=2$) gives $R_o=1.7\times10^{16}$~cm and $xR_o=3.2\times 10^{14}$~cm
(E1, E2).
For $w=0$, $R_o-xR_o\simeq v_st_d=6\times10^{15}$~cm (c.f. $y_1R_o\simeq
R_o$).
As $y_1R_o=1.1\times10^{16}$~cm in $w=5$ model, we adopted
$R_o=1.1\times 10^{16}$~cm and $xR_o=R_i=5\times 10^{15}$~cm (F1).

\section{Light Curve Models}\label{LC}
Starting from the initial conditions obtained in Section \ref{sec2},
we perform the numerical LC calculations with \verb|STELLA|.
We show that the $v_s=5,200~\mathrm{km~s^{-1}}$ models cannot
explain the huge luminosity of SN 2006gy self-consistently.
LCs obtained from $v_s=10,000~\mathrm{km~s^{-1}}$ models
are broadly consistent with the observational properties of
SN 2006gy.

\subsection{$v_s=5,200~\mathrm{km~s^{-1}}$ Model}\label{sec:05}
\begin{figure}
\begin{center}
 \includegraphics[width=\columnwidth]{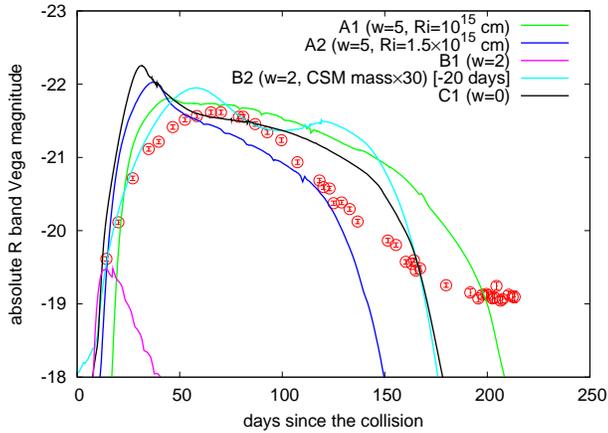}
  \caption{
Absolute $R$ band LCs of the $v_s=5,200~\mathrm{km~s^{-1}}$ models.
These models are not self-consistent.
The origin of time axis is set to when our numerical calculations start,
i.e., when the SN ejecta and CSM start to collide, except for B2.
The time of the model B2 is shifted $-20$ days.
}
\label{RLC52}
\end{center}
\end{figure}

Figure \ref{RLC52} shows the LCs from the $v_s=5,200~\mathrm{km~s^{-1}}$
models with the observed $R$ band LC of \citet{smith2007b}.
$E_\mathrm{ej}$ is chosen so that the peak luminosities of the model
LCs can be as luminous as that of SN 2006gy.
However, $E_\mathrm{ej}$ should be
very high $(\simeq 5\times 10^{52}~\mathrm{erg})$
for the $v_s=5,200~\mathrm{km~s^{-1}}$ models to be as luminous as SN 2006gy
and assuming the relatively low $v_s=5,200~\mathrm{km~s^{-1}}$
is not consistent with the high kinetic energy.
This inconsistency can also be seen from the rising times of the models.
Although the models are constructed so that the rising times of the LCs
become $t_d=70$ days, the rising times of the numerical results
are much shorter than 70 days.
If we set smaller $E_\mathrm{ej}$, the rising times can be
the same as that of SN 2006gy but then the luminosities
become much smaller than that of SN 2006gy.
Shortly, the models derived by assuming 
$v_s=5,200~\mathrm{km~s^{-1}}$ are not compatible with
the large luminosity of SN 2006gy.
Given these results, we adopt models with a higher $v_s$,
$v_s=10,000~\mathrm{km~s^{-1}}$, and
they are found to be able to explain the LC of SN 2006gy
self-consistently (Section \ref{sec:10}).
%The $v_s=10,000~\mathrm{km~s^{-1}}$
%models are found to be also consistent with the evolution
%of the blackbody radius of SN 2006gy based on which the shock velocity
%$v_s=5,200~\mathrm{km~s^{-1}}$ is estimated (Section \ref{sec:dynamical}).

One important question of the interaction model is
whether the CSM from the steady mass loss $(w=2)$ can explain
the properties of SLSNe and we look into the $w=2$ models more carefully.
If it can, a mechanism to achieve such huge steady mass loss may exist.
If not, it is indicated that
explosive non-steady mass loss takes place in their progenitors
and there should exist some mechanisms to cause such mass loss
just before their explosions.

The $w=2$ model B1 reaches only $\simeq -19.5$ mag in the $R$ band at the LC peak.
This is because $M_\mathrm{CSM}=0.83 M_\odot$ in the B1 model is
much smaller than $M_\mathrm{CSM}$ of the models with the other $w$.
The fraction of kinetic energy converted to radiation in the model B1 is 
much smaller than those in the models A1, A2, and C1,
as the amount of energy converted from kinetic energy to radiation
strongly depends on the relative mass of CSM and SN ejecta (see Section \ref{sec:eff}).
Thus, more kinetic energy is required 
for the B1 model to be as luminous as SN 2006gy.
However, the rising time of the B1 model is already much less than
that of SN 2006gy and it becomes shorter if we increase kinetic energy.
Thus, the $v_s=5,200~\mathrm{km~s^{-1}}$ model with the steady wind
($w=2$, B1) is hard to be compatible with SN 2006gy.
For demonstration, we also calculated a model (B2) in which $M_\mathrm{CSM}$
is increased 30 times more than that of the model B1.
Then, the amount of energy converted increases because of 
the high efficiency for the energy conversion.
In addition, the photospheric radius is increased due to the increased density.
As a result, the luminosity of the model becomes as large as that of SN 2006gy.

\begin{figure}
\begin{center}
 \includegraphics[width=\columnwidth]{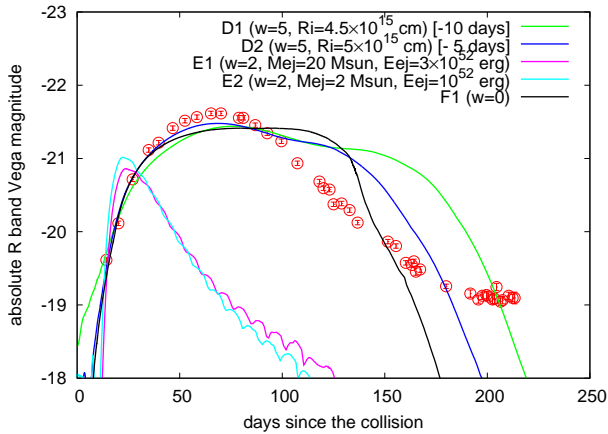}
  \caption{
Absolute $R$ band LCs of the $v_s=10,000~\mathrm{km~s^{-1}}$ models.
These models can self-consistently explain the rising time and the peak
 luminosity of SN 2006gy.
The origin of the time axis is $+10$ days (D1), $+5$ days (D2),
$0$ days (E1, E2, and F1) since the collision.
}
\label{RLC100}
\end{center}
\end{figure}

\subsection{$v_s=10,000~\mathrm{km~s^{-1}}$ Model}\label{sec:10}
As $v_s=5,200~\mathrm{km~s^{-1}}$ models are not
able to explain SN 2006gy self-consistently,
we investigate models with higher $v_s$, $v_s=10,000~\mathrm{km~s^{-1}}$.
$v_s=5,200~\mathrm{km~s^{-1}}$ is estimated from
the evolution of the blackbody radius but it is shown that
the evolution of the blackbody radius in the
$v_s=10,000~\mathrm{km~s^{-1}}$ models is consistent
with that of SN 2006gy.

\subsubsection{Light Curve}
\begin{figure*}
\begin{center}
 \includegraphics[width=\columnwidth]{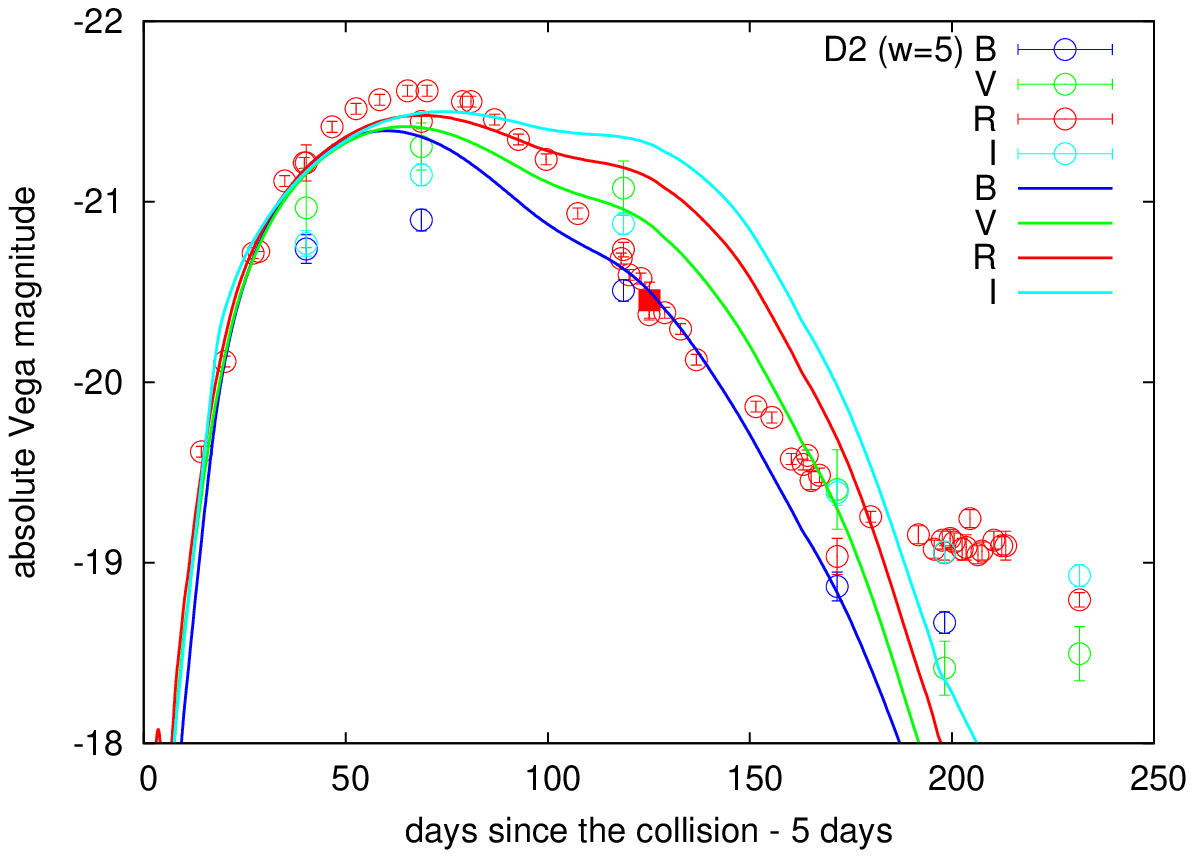}
 \includegraphics[width=\columnwidth]{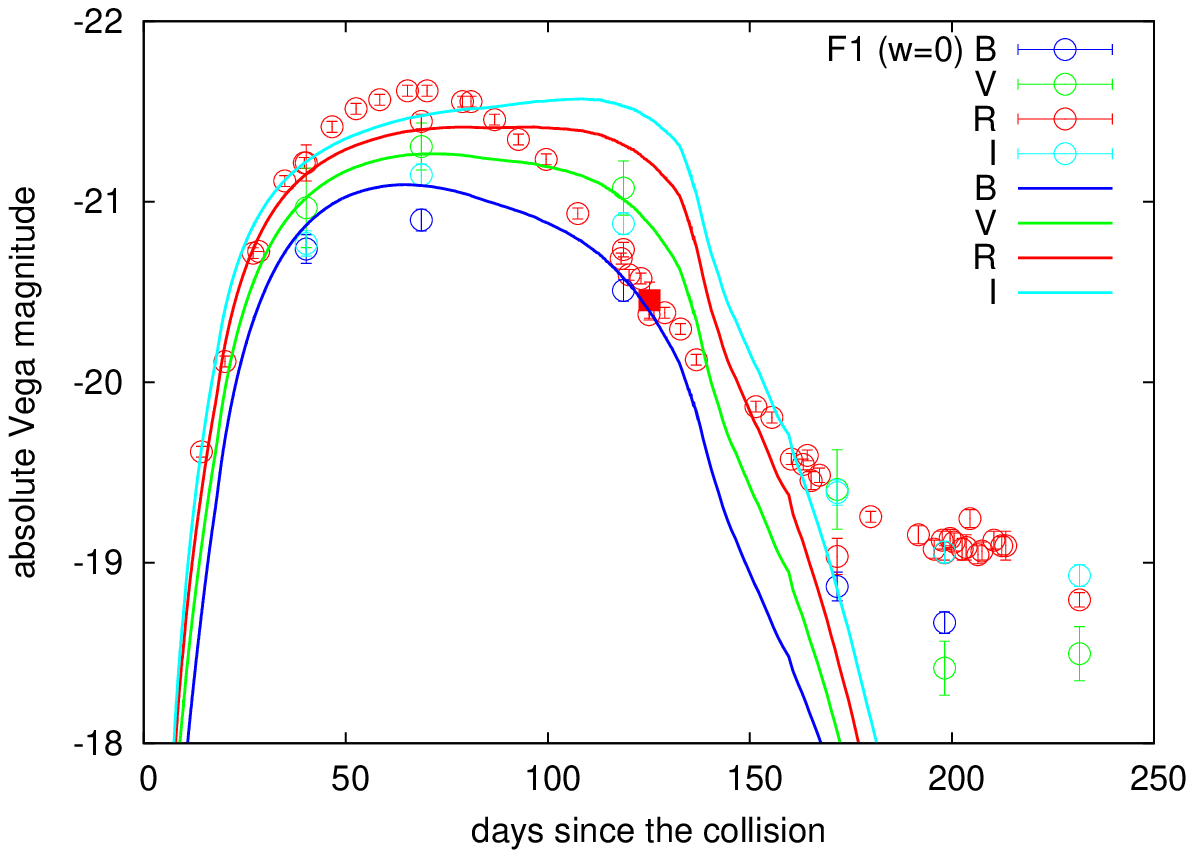}
  \caption{
Multicolor LCs of the models D2 ($w=5$, left) and F1 ($w=0$, right).
The observational data are from
\citet{smith2007b,agnoletto2009,kawabata2009}.
The origin of the time axis
in the left (right) panel is 5 (0) days since the collision.
}
\label{multi}
\end{center}
\end{figure*}

\begin{figure}
\begin{center}
 \includegraphics[width=\columnwidth]{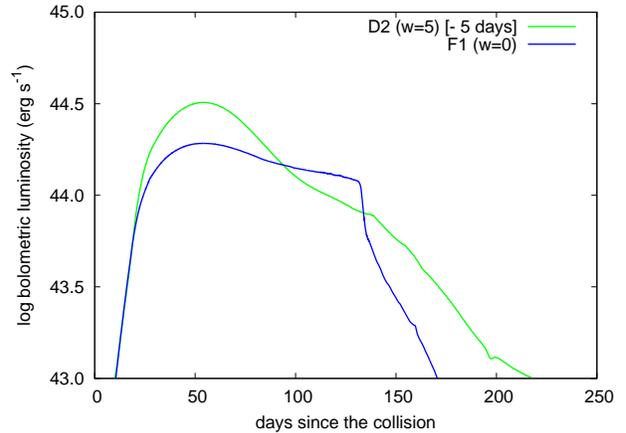}
  \caption{
Bolometric LCs of the models D2 ($w=5$) and F1 ($w=0$).
The origin of the time axis
in the D2 (F1) model is 5 (0) days since the collision.
}
\label{bol}
\end{center}
\end{figure}

\begin{figure*}
\begin{center}
 \includegraphics[width=\columnwidth]{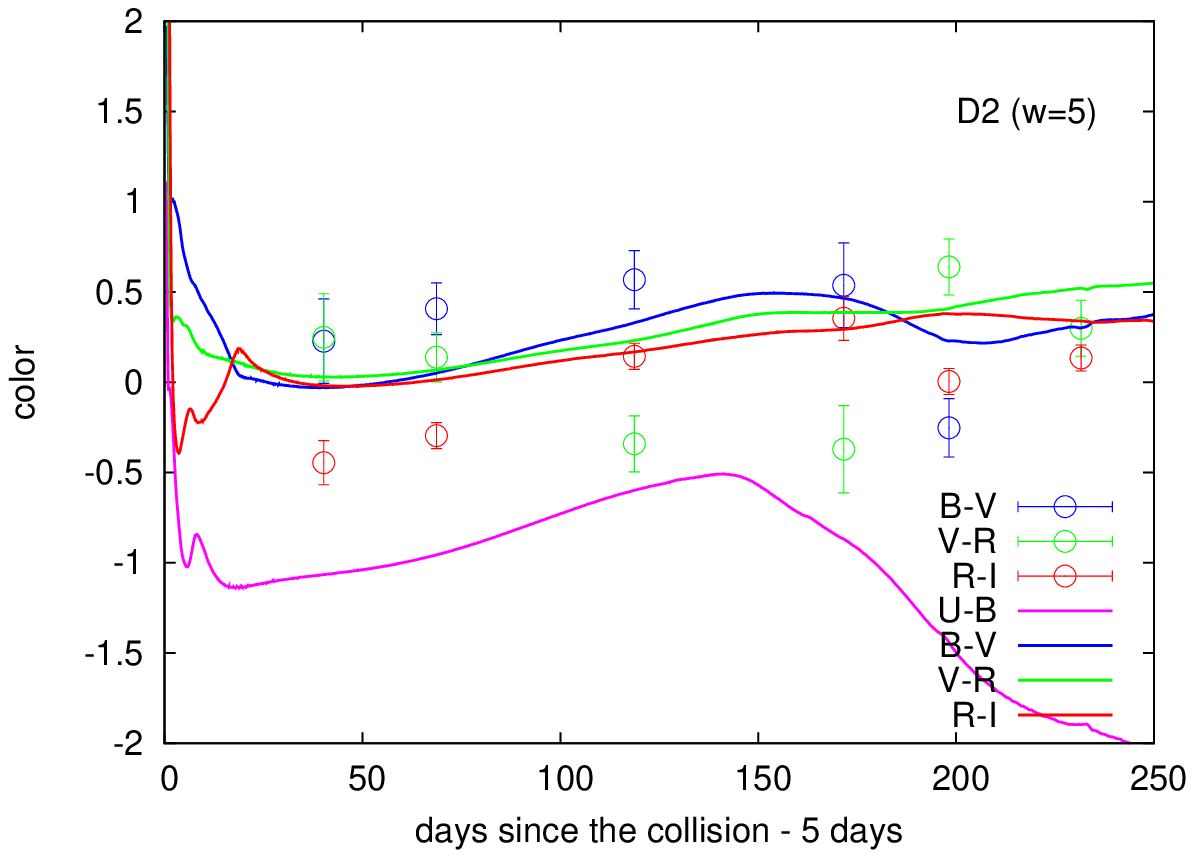}
 \includegraphics[width=\columnwidth]{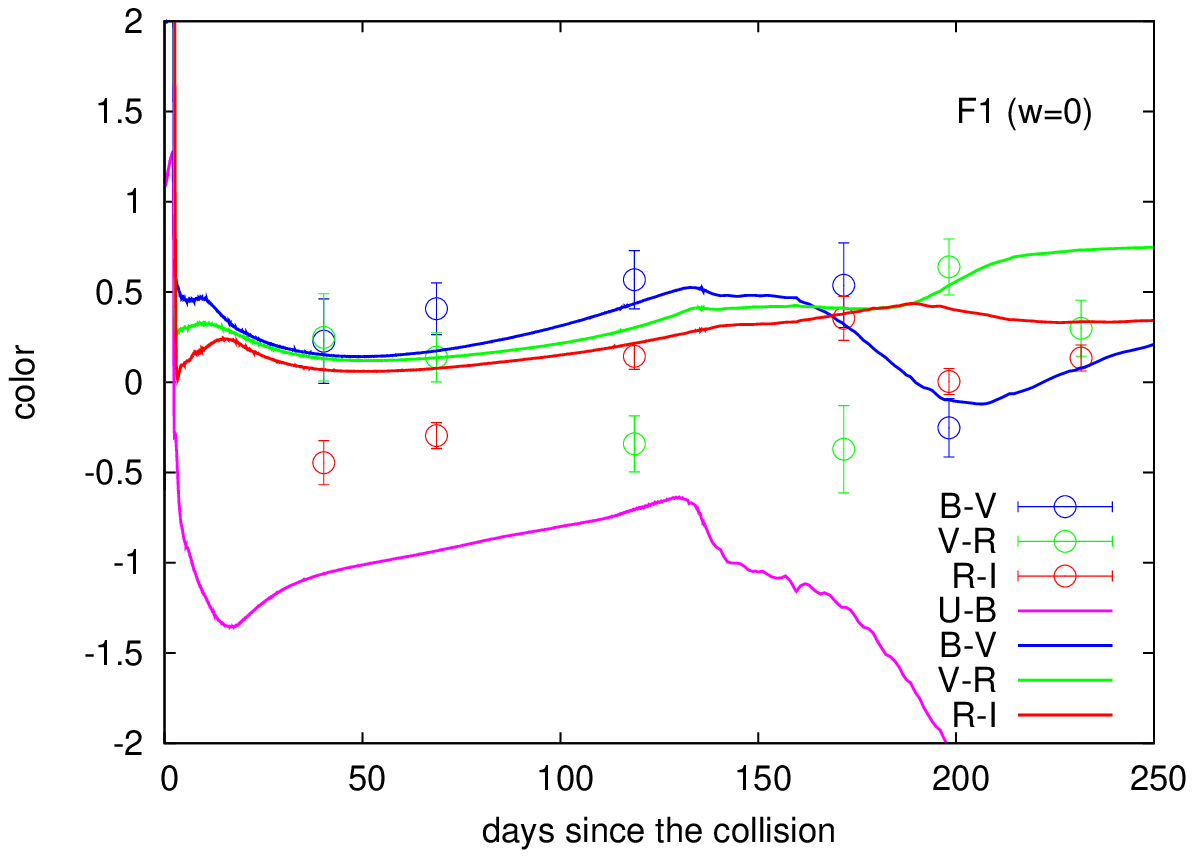}
  \caption{
Color evolution of the models D2 ($w=5$, left) and F1 ($w=0$, right).
Observational points are from \citet{agnoletto2009}.
}
\label{color}
\end{center}
\end{figure*}

The $R$ band LCs from the $v_s=10,000~\mathrm{km~s^{-1}}$ models are
shown in Figure \ref{RLC100}.
Multicolor LCs of the models D2 $(w=5)$ and F1 $(w=0)$ are shown in Figure
\ref{multi} and the bolometric
LCs of the two models are shown in Figure \ref{bol}.
The color evolution of the models D2 and F1 are shown in Figure \ref{color}.
%SEDs of the models D2 and F1 at some selected epochs are shown in
%Figure \ref{sed}.

The rising parts and the peak luminosities of the LCs of the $w=0,5$ models are
consistent with SN 2006gy.
Thus, the $v_s=10,000~\mathrm{km~s^{-1}}$ models are
self-consistent with the assumed $E_\mathrm{ej}$ and $M_\mathrm{ej}$.
The $v_s=10,000~\mathrm{km~s^{-1}}$ models only require
$E_\mathrm{ej}=10^{52}$ erg to achieve the peak luminosity of SN 2006gy,
instead of $E_\mathrm{ej}\simeq 5\times 10^{52}$ erg
required for the $v_s=5,200~\mathrm{km~s^{-1}}$ models.
This is because the blackbody radius
in the CSM can be larger in the $v_s=10,000~\mathrm{km~s^{-1}}$ models
and less energy is required to achieve the same luminosity.
The steady mass-loss models $(w=2)$ are, however, still not consistent with
SN 2006gy.

The model LCs with $w=0,5$ after the peak
start to deviate from the observed LCs, although the deviations
stay less than one magnitude 
 before the plateau in the observed LC at around 200 days.
Our model $R$ band LCs take some time after the LCs have reached the peak
until the LCs start to decline,
contrary to the observed $R$ band LC of SN 2006gy.
This is because there remains unshocked optically thick CSM
even after the LC peak in our numerical models and
the photosphere remains there for a while.
The analytic model of \citet{moriya2011}
which we use for the estimate of the initial condition
assumes a constant $v_s$.
However, $v_s$ actually reduces as the interaction goes on
and the optically thick part of CSM does not
shocked away entirely at the time $t_d$ when the optically thick CSM
is assumed to be swept up by the forward shock 
in the model of \citet{moriya2011}.
This effect is more significant in $w=0$ models
because $w=0$ models suffer more on the deceleration than
$w=5$ models.
A severe failure of our models is that all of them fail to
reproduce the plateau in the LC of SN 2006gy at around 200 days.
We discuss this separately in Section \ref{plateau}.

Looking at the multicolor LCs (Figure \ref{multi}), the $w=0$ model
(F1) is closer to the observed LC, especially the 
$B$ band LC of the rising epochs.
This is presumed to be because the initial density jump between SN ejecta and CSM is
smaller in the $w=0$ model (Figure \ref{density})
and the temperature becomes lower 
in the $w=0$ model. However, the LCs in the $U$ and $B$ bands can be
affected by many weak absorption lines of Fe group elements
which are not taken into account in our opacity.
Those weak absorptions may reduce the luminosity of the $U$ and $B$ band
LCs and we cannot judge which model is better based just on the 
blue part of the LCs.
In addition, the difference is $\simeq 0.5$ mag
and they are not significant.
The color evolution of the two models (Figure \ref{color})
roughly follows the observed evolution,
although there exist some deviations especially in $R-I$ and $V-R$.

Although it is possible to continue LC modeling to get much
better fits to the SN 2006gy LC, it does not necessarily lead us
to the better understanding of the properties
of the SN ejecta and the dense CSM involved in the progenitor system of
SN 2006gy.
This is mainly because of the simplified physics adopted in \verb|STELLA|.
Especially, \verb|STELLA| is a one-dimensional code and 
multidimensional effects are approximately incorporated by adopting the smearing
parameter. As the uncertainties involved in the parameter are
large (see Section \ref{sm}), making a perfect fit to the observed
LC does not necessarily provide us with the best parameter.
In addition, 
the differences in the LCs in the declining phases are
less than one magnitude (or a factor $\simeq 2$)
and the differences in the rising phases
are much less. Thus, the properties of the SN ejecta and the dense CSM
in the D2 and F1 models are presumed not to be so different from the 'actual'
values. 
Thus, we conclude that the CSM parameters predicted by the shock
breakout model can explain the overall properties of SN 2006gy.
We also note that
a systematical study of the effect of the CSM properties on the
LCs powered by the interaction between SN ejecta and dense CSM
is summarized in \citet{moriya2011b}.
The durations of the LCs of the models D2 and F1 are a bit longer than
that of SN 2006gy.
To reduce the durations of the LCs by keeping the peak luminosities of them,
we can, for example, change the radii of the CSM.

Finally, we look into the steady mass loss models
to see whether they actually fail to reproduce the LC of SN 2006gy.
The steady mass-loss model ($w=2$, E1) is, again, too faint
to explain SN 2006gy with $E_\mathrm{ej}=3\times 10^{52}$ erg.
The rising time is already too short and
reaching the peak luminosity of SN 2006gy by
increasing $E_\mathrm{ej}$ does not work
as is discussed in Section \ref{sec:05}.
This is because of the too small $M_\mathrm{CSM}$ and
can be improved if $M_\mathrm{CSM}$ is increased (see the models B1 and B2).
Another possible way to make $w=2$ models work is to increase
the conversion efficiency from $E_\mathrm{ej}$ to radiation energy
so that $E_\mathrm{ej}$ can be reduced (see Section \ref{sec:eff} for the
discussion of the conversion efficiency).
In the model E2, $M_\mathrm{ej}$ is set to be comparable to
$M_\mathrm{CSM}$ so that the conversion efficiency
becomes higher (Section \ref{sec:eff}).
The similar peak luminosity to the model E1 is reached
with less $E_\mathrm{ej}$ $(10^{52}~\mathrm{erg})$ in the model E2.
However, as the diffusion time of the CSM is not affected so much by this,
the LCs become similar to each other
and increasing the efficiency does not revive the $w=2$ models.
To summarize, the dense CSM from the steady mass loss
is still difficult to explain the LC of SN 2006gy with the shock
breakout model.

%\begin{figure*}
%\begin{center}
% \includegraphics[width=\columnwidth]{d2sed.eps}
% \includegraphics[width=\columnwidth]{f1sed.eps}
%  \caption{
%SEDs of the models D2 ($w=5$, left) and F1 ($w=0$, right).
%Days since the LC peak are shown in the figure.
%}
%\label{sed}
%\end{center}
%\end{figure*}
\begin{figure*}
\begin{center}
 \includegraphics[width=\columnwidth]{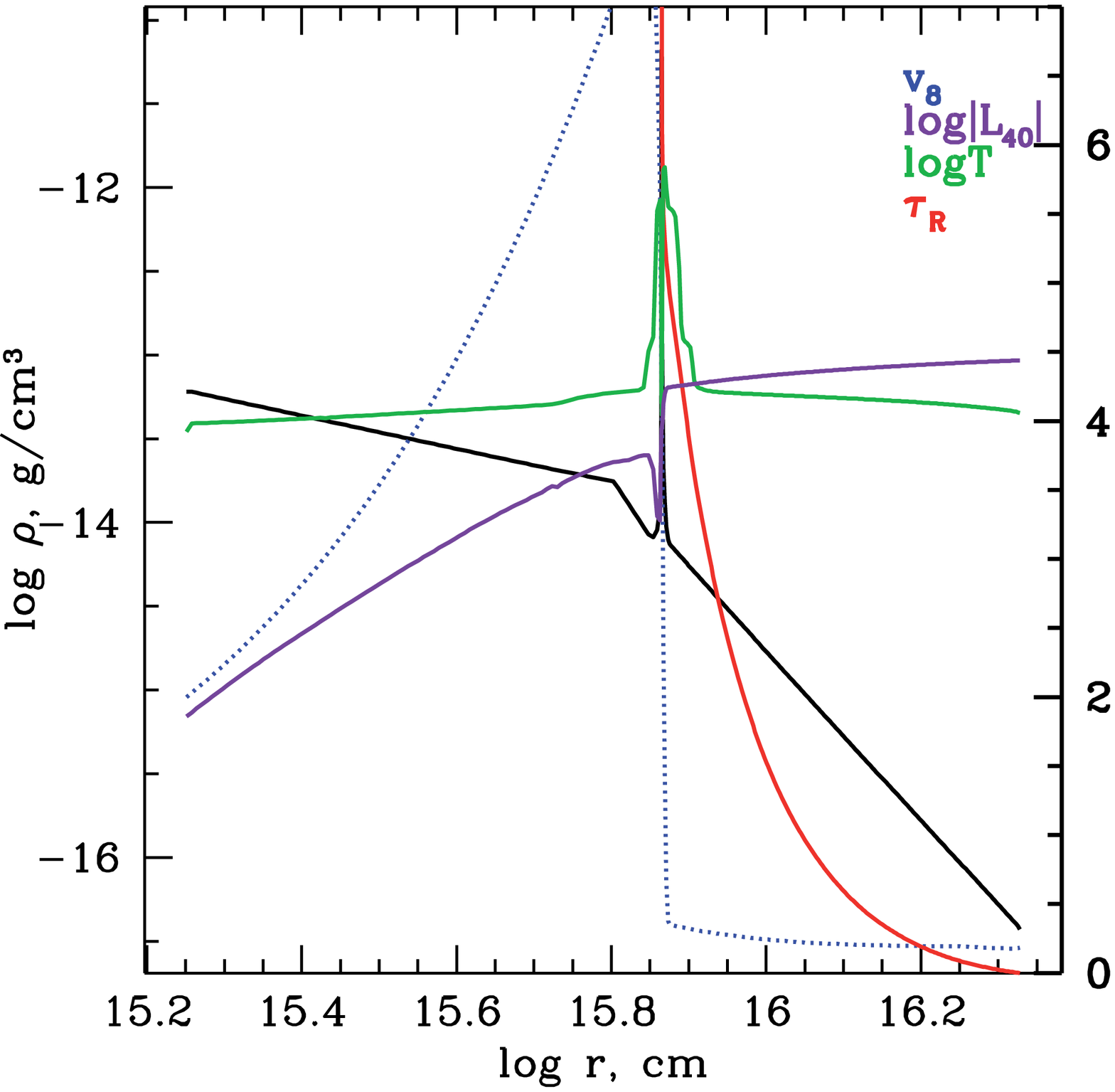}
 \includegraphics[width=\columnwidth]{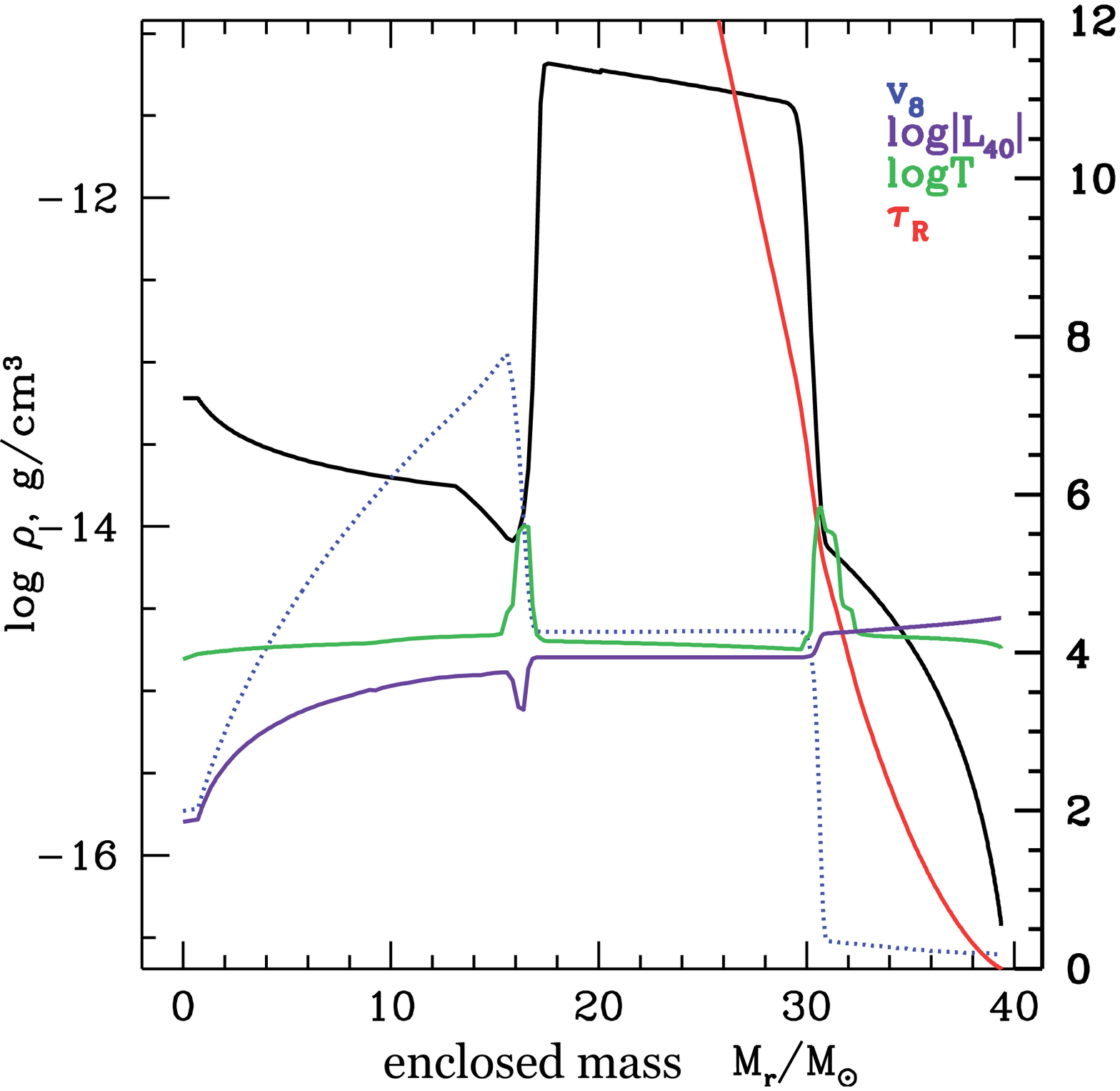}
  \caption{
Physical structures of the model D2 in
radius (left) and mass coordinate (right)
at around the LC peak (74 days since the collision).
Black lines show the density structure (left $y$-axis).
Blue dotted lines are the velocity scaled by $10^{8}~\mathrm{cm~s^{-1}}$ (right
$y$-axis), purple lines are the logarithm of the
absolute value of luminosity scaled by $10^{40}~\mathrm{erg~s^{-1}}$
 (right $y$-axis),
green lines are the logarithm of the temperature in Kelvin (right
 $y$-axis), and
red lines are Rosseland optical depth measured from the outside
(right $y$-axis).
}
\label{rm}
\end{center}
\end{figure*}

\subsubsection{Dynamical Evolution}\label{sec:dynamical}
Figure \ref{rm} shows the dynamical structures of the model D2 $(w=5)$ at around
the LC peak. The left panel shows the structure in the physical
coordinate (radius) whereas
the right panel shows the structure in the mass coordinate.
The cool dense shell is created between SN ejecta and CSM in which
about $10~M_\odot$ of shocked CSM and $3~M_\odot$ of shocked SN ejecta are contained.

\begin{figure}
\begin{center}
 \includegraphics[width=\columnwidth]{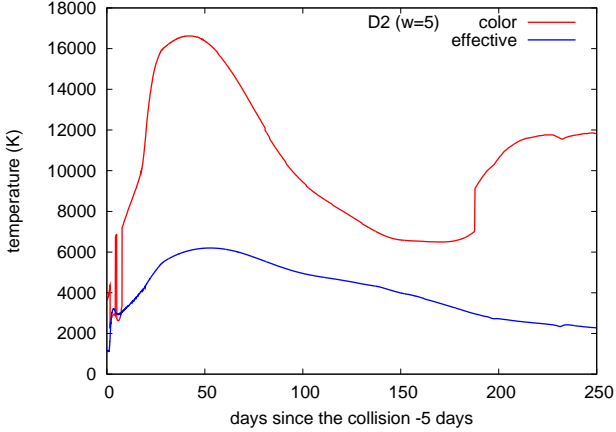}
  \caption{
Evolution of the color temperature $(T_\mathrm{col})$
and the effective temperature $(T_\mathrm{eff})$
in the model D2.
The origin of the time axis is 5 days since the collision.
}
\label{temperature}
\end{center}
\end{figure}

\begin{figure}
\begin{center}
 \includegraphics[width=\columnwidth]{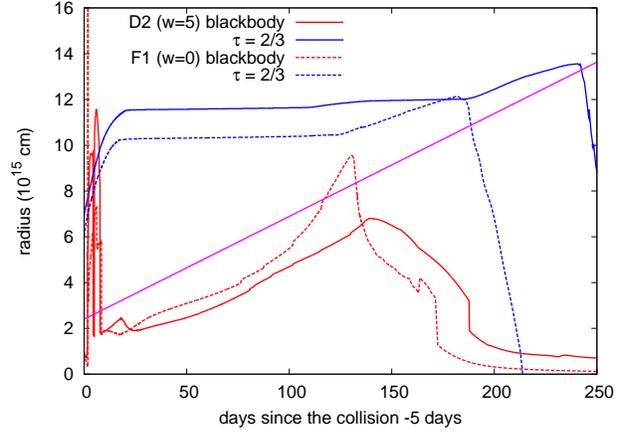}
  \caption{
Evolution of the blackbody radius ($R_\mathrm{BB}$, solid line)
and the photospheric radius ($R_{\tau=2/3}$, dashed line) of
the models D2 ($w=5$) and F1 ($w=0$).
The origin of the time axis is 5 days since the collision.
The monotonically increasing linear line at the middle is
the evolution of the blackbody radius obtained by \citet{smith2010},
a constant velocity evolution with $5,200~\mathrm{km~s^{-1}}$.
The observational blackbody radius follows the line
until around 125 days and starts to decline
(see \citet{smith2010} for details).
}
\label{bbradius}
\end{center}
\end{figure}

Figure \ref{temperature} shows the evolution of the color temperature $(T_\mathrm{col})$
and the effective temperature $(T_\mathrm{eff})$ of the model D2.
The color temperature is derived by fitting the spectra
obtained by the numerical calculations
with the blackbody spectral distribution
whereas the effective temperature is obtained by
using the bolometric luminosity $(L_\mathrm{bol})$ and
the radius $(R_{\tau_R=2/3})$ of the photosphere
which is defined as the radius where the
Rosseland mean optical depth $\tau_R$ from the surface becomes
$2/3$ in \verb|STELLA| and is expressed as
$T_\mathrm{eff}=\left(L_\mathrm{bol}/4\pi\sigma R_\mathrm{\tau_R=2/3}^2\right)^{1/4}$.
Here, $\sigma$ is the Stephan-Boltzmann constant.
As radiation mainly comes from the shell and
the Thomson scattering is the dominant opacity source in the CSM above the shell,
$T_\mathrm{col}$ roughly traces the temperature of the shell.
The photosphere ($R_{\tau_R=2/3}$)
is much above the shell and $T_\mathrm{eff}$
becomes very low because of the large $R_{\tau_R=2/3}$ (see also Figure \ref{bbradius}).
At the time when $T_\mathrm{col}$ starts to increase for the second time
(from $\simeq 180$ days),
the photosphere is in the SN ejecta whose density structure and
composition are expressed in the approximated way and the results around
these epochs and later should not be taken seriously.

Figure \ref{bbradius} shows the evolution of the blackbody radius
$R_\mathrm{BB}=\sqrt{L_\mathrm{bol}/4\pi \sigma T_\mathrm{col}^4}$
and the photosphere $R_{\tau_R=2/3}$.
The constant velocity line with $5,200~\mathrm{km~s^{-1}}$
is the evolution of the blackbody radius obtained by \citet{smith2010}
along which $R_\mathrm{BB}$ follows until around 125 days.
The evolution of $R_\mathrm{BB}$ in the models D2 and F1 is consistent with
$5,200~\mathrm{km~s^{-1}}$ although the radius is a bit smaller than
the observed values.
%Note that $R_\mathrm{BB}$ before $\sim 50$ days is consistent with
%the high-temperature interpolation of observations
%\citep[see Figure 7 of][]{smith2010}.
The bolometric correction of \citet{smith2010}
is based on $T_\mathrm{BB}$ and
the correction may add extra luminosities
because it ignores the effect of line depletion.
Higher $L_\mathrm{bol}$ results in higher $R_\mathrm{BB}$
for a given $T_\mathrm{col}$ and this can be the reason
why $R_\mathrm{BB}$ of \citet{smith2010} is higher than ours.

$R_\mathrm{BB}$ obtained by our calculations tends to be smaller than
the shell radius where the radiation is coming from.
For example, $R_\mathrm{BB}$ of the model D2 shown in
Figure \ref{bbradius} stays lower than the radius
at which the interaction starts $(R_i=5\times 10^{15}~\mathrm{cm})$.
The reason is presumed to be similar to that of the discrepancy
in $R_\mathrm{BB}$ obtained from observations and numerical calculations.
$T_\mathrm{col}$ is obtained from the spectral fitting but
actual spectra suffer from the line depletion especially
in blue.
Since $T_\mathrm{col}$ is reduced from the temperature
at the photon production site,
$L_\mathrm{bol}$ is the value affected by such depletion
and is less than the value expected from the blackbody with $T_\mathrm{col}$.
Thus, with the smaller $L_\mathrm{bol}$,
$R_\mathrm{BB}=\sqrt{L_\mathrm{bol}/4\pi \sigma T_\mathrm{col}^4}$
becomes smaller than the actual emitting region.

As most of hydrogen in CSM remains to be ionized, $R_{\tau_R=2/3}$
continues to be at the radius where
the Rosseland mean opacity from the surface of the CSM is $2/3$ and remains
to be constant until the shock wave comes close to the radius.
Then, $R_{\tau_R=2/3}$ evolves roughly following the forward shock.

\subsubsection{$M_\mathrm{ej}$ and $E_\mathrm{ej}$}\label{sec:mejeej}
\begin{figure}
\begin{center}
 \includegraphics[width=\columnwidth]{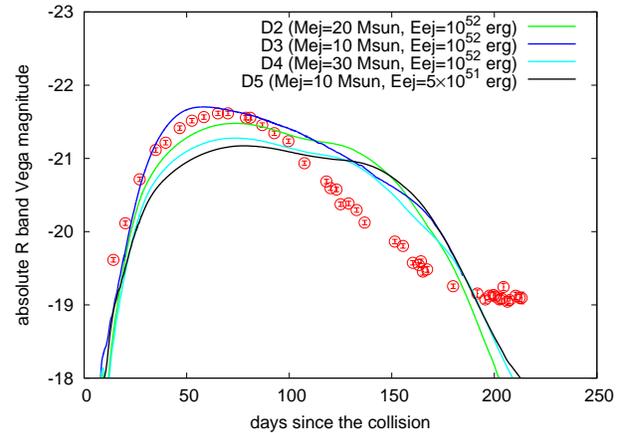}
  \caption{
$R$ band LCs from the same CSM but different $M_\mathrm{ej}$ and
$E_\mathrm{ej}$.
The models D2, D3, D4 have the same $E_\mathrm{ej}$ but different
$M_\mathrm{ej}$, i.e., $10~M_\odot$ (D3), $20~M_\odot$ (D2), and
 $30~M_\odot$ (D3).
The model D5 has less $E_\mathrm{ej}\ (5\times 10^{51}~\mathrm{erg})$
than other models $( 10^{52}~\mathrm{erg})$ and $M_\mathrm{ej}=10~M_\odot$.
The calculation of the LC of the model D3 stopped at around 175 days
since the collision and we show the LC until around 175 days.
}
\label{mejekin}
\end{center}
\end{figure}

The properties of SN ejecta ($M_\mathrm{ej}$ and $E_\mathrm{ej}$)
determine many aspects of SNe powered by
the shock interaction (e.g., luminosities)
because $E_\mathrm{ej}$ determines the available energy and
$M_\mathrm{ej}$ affects the efficiency to convert the available
kinetic energy to the radiation energy.
We discuss the effect of $M_\mathrm{ej}$ and $E_\mathrm{ej}$
in Section \ref{sec:eff} including the results
of the $v_s=5,200~\mathrm{km~s^{-1}}$ models and here we just show the
results of LC calculations with different $M_\mathrm{ej}$ and
$E_\mathrm{ej}$ (Figure \ref{mejekin}).
LCs are similar to each others and we can see that it is difficult to
constrain $M_\mathrm{ej}$ and $E_\mathrm{ej}$ only by the LC.
This can also be seen by comparing the models E1 and E2 in Figure \ref{RLC100}.
The two models have different $M_\mathrm{ej}$ and $E_\mathrm{ej}$
with the same CSM but the resulting LCs are similar
(see discussion in Section \ref{sec:eff}).

In the best LC model of the pulsational pair-instability model
presented by \citet{woosley2007}, the ejecta with $5.1~M_\odot$
and $2.9\times 10^{51}$~erg collides the CSM with $24.5~M_\odot$.
Our canonical models (D2 and F1) have much higher $E_\mathrm{ej}$
($10^{52}$ erg).
One of the reasons is presumed to be the smaller photospheric radius
($y_1R_o=1.1\times 10^{16}$ cm) in our models.
The dense CSM in the pulsational pair-instability model extends to
about $3\times 10^{16}$ cm and the photosphere can be larger than
our models. This effect of the locations of the photosphere can also be seen
in comparison to the $v_s=5,200~\mathrm{km~s^{-1}}$ models.
The photospheric radii of them are only $y_1R_o\simeq 5\times 10^{15}$~cm
at most
and the required energy to achieve the maximum luminosity of SN 2006gy
is $5\times 10^{52}$~erg which is even larger than
$10^{52}$~erg required for the $v_s=10,000~\mathrm{km~s^{-1}}$ models
($y_1R_o=1.1\times10^{16}$~cm). 
This shows the difficulties to constrain
$E_\mathrm{ej}$ only by the LC.

In addition, the efficiency to convert the kinetic
energy to radiation is mainly determined by the relative mass of the ejecta and
the collided CSM. It does not depend strongly
on the ejecta mass if the CSM mass is much larger than the ejecta mass.
To get high conversion efficiencies of the
kinetic energy to the radiation energy, $M_\mathrm{ej}$ is better
to be comparable or less than $M_\mathrm{CSM}$
and we can at least get some constraint on $M_\mathrm{ej}$
from the LC based on the view point of the conversion
efficiency (Section \ref{progenitor}).

\subsubsection{Smearing}\label{sm}
The dense shell which appears between SN ejecta and CSM is unstable
in multidimension as is discussed in Section \ref{sec2:sm}.
As a result of the instabilities, less kinetic energy is expected to be converted to radiation
because there would be the extra multidimensional motions caused by the instabilities.
To take into account such multidimensional effects in one-dimensional code
\verb|STELLA|, we include a smearing term in the equation of motion
(the parameter $B_q$, Section \ref{sec2:sm}).
%The multidimensional effects reduce the efficiency
%to convert the kinetic energy of SN ejecta to radiation energy,
%in other words, the cooling efficiency in the shell is reduced.

\begin{figure}
\begin{center}
 \includegraphics[width=\columnwidth]{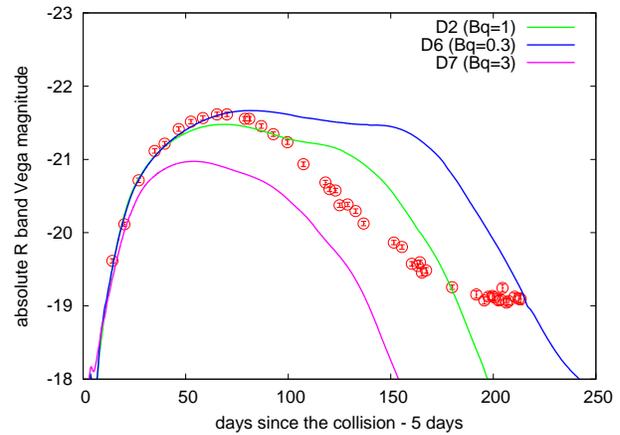}
  \caption{
$R$ band LCs with different $B_q$, i.e.,
D2 ($B_q=1$), D6 ($B_q=0.3$), and D7 ($B_q=3$).
The smearing parameter $B_q$ changes the conversion efficiency
from kinetic energy to radiation.
With Larger $B_q$, more kinetic energy remains 
and less radiation energy is emitted.
The origin of the time axis is 5 days since the collision.
}
\label{BQ}
\end{center}
\end{figure}

Figure \ref{BQ} shows the LCs with different values of the
smearing parameter $B_q$.
With larger $B_q$, the effect of the smearing becomes larger
and less kinetic energy is converted to radiation.
In other words, radiative cooling becomes less efficient.
The model D2 is calculated with our standard $B_q=1$.
The model D6 has $B_q=0.33$ and the model D7 has $B_q=3$.
The shape of the LC is different even if we only change
$B_q$ with a factor 3.
We also show the effect of $B_q$ on
the LCs obtained from the pulsational pair-instability
SN models of \citet{woosley2007} in Appendix \ref{appendix}
and show that the effect is not unique to our models.
We discuss the efficiency in detail in Section \ref{sec:eff}.

The uncertainty in the smearing parameter adds another
difficulty in our estimations of physical parameters of the
progenitor system. This is one reason why we think that
making the perfect fitting now does not lead us to the
exact parameters of the progenitor system.
The calibrations for the smearing parameter should be done at least.
However, the rising time and the peak luminosity is not
so sensitive to the smearing parameter and the parameters
of SN ejecta and dense CSM we obtain
with the current uncertainty are presumed to be close to
the real ones.

\subsubsection{Effect of {\rm \Ni}}\label{sec:ni}
\begin{figure}
\begin{center}
 \includegraphics[width=\columnwidth]{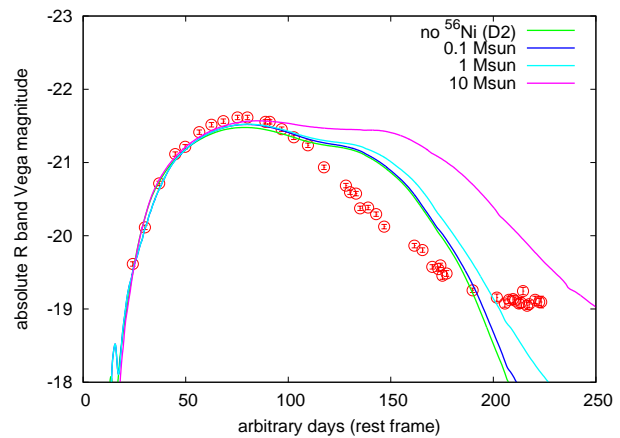}
  \caption{
$R$ band LCs with \Ni.
$0.1~M_\odot$, $1~M_\odot$, and $10~M_\odot$ of
\Ni\ is included at the center of the model D2.
Only $10~M_\odot$ of \Ni\ makes significant effect on the LC.
}
\label{ni}
\end{center}
\end{figure}

We have also examined the effect of \Ni\ decay on the LCs.
Figure \ref{ni} shows the results.
We include \Ni\ at the center of the model D2.
If we include \Ni, the length of the peak is extended
due to the extra heat source.
The significant effect can only be seen when we
include $\sim 10~M_\odot$ of \Ni. However, the amount of \Ni\
is observationally constrained to be less than $2.5~M_\odot$
\citep{miller2010} and the effect of \Ni\ is negligible.

\section{Discussion}\label{discussion}
\subsection{Conversion Efficiency}\label{sec:eff}
The source of radiation in our LC models is the kinetic energy
of SN ejecta.
The amount of energy converted from kinetic energy to radiation
can be estimated by the conservation laws of energy and momentum.
If we assume that the radiation pressure
does not change the dynamics of the
materials so much, the conservation of momentum requires
\begin{eqnarray}
M_\mathrm{colej}v_\mathrm{colej}&=&\left(M_\mathrm{colej}+M_\mathrm{sCSM}\right)
v_\mathrm{shell},\label{ef2}
\end{eqnarray}
where $M_\mathrm{colej}$ is the mass of the collided
SN ejecta, $v_\mathrm{colej}$ is the mean velocity of
the collided SN ejecta, $M_\mathrm{sCSM}$ is the mass
of the shocked CSM, and $v_\mathrm{shell}$ is the velocity of
the dense shell between SN ejecta and CSM.
Radiation energy $E_\mathrm{rad}$ emitted as a result of
the interaction can be derived from the conservation of energy
\begin{eqnarray}
E_\mathrm{rad}&=&
\alpha\left[
\frac{1}{2}M_\mathrm{colej}v_\mathrm{colej}^2
-\frac{1}{2}\left(M_\mathrm{colej}+M_\mathrm{sCSM}\right)v_\mathrm{shell}^2
\right], \label{ef3}
\end{eqnarray}
where $\alpha$ is the fraction of kinetic energy converted to radiation.
From Equations (\ref{ef2}) and (\ref{ef3}),
\begin{equation}
\frac{E_\mathrm{rad}}{\frac{1}{2}M_\mathrm{colej}v_\mathrm{colej}^2}
=\frac{\alpha M_\mathrm{sCSM}}{M_\mathrm{colej}+M_\mathrm{sCSM}}.\label{bare}
\end{equation}
If most of the SN ejecta and CSM are shocked,
i.e., $M_\mathrm{colej}\simeq M_\mathrm{ej}$ and $M_\mathrm{sCSM}\simeq M_\mathrm{CSM}$,
we get the rough estimate for the radiation energy emitted
\begin{equation}
E_\mathrm{rad}\simeq\frac{\alpha M_\mathrm{CSM}}{M_\mathrm{ej}+M_\mathrm{CSM}}E_\mathrm{ej}.
\label{ef4}
\end{equation}
$\alpha$ is expected to be close to 1 without the smearing parameter $B_q$
because most of thermal energy gained by the shock
is eventually emitted as radiation.
Since the parameter $B_q$ adds additional acceleration 
to reduce the amount of energy converted to thermal energy,
$\alpha$ is expected to become lower as $B_q$ becomes larger.
The rest of energy is mostly in the form of kinetic energy.
We may also express the effect as the reduction of the radiative cooling
efficiency because less radiation energy is emitted
with the smearing term.

\begin{figure}
\begin{center}
 \includegraphics[width=\columnwidth]{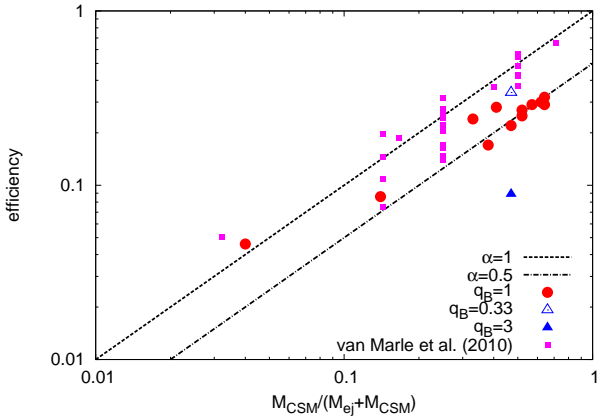}
  \caption{
Efficiency of the conversion of kinetic energy to radiation.
Most of the results from \citet{vanmarle2010}, including
aspherical models, are also shown.
$\alpha$ is the measure for the effect of the smearing parameter $B_q$.
$\alpha=1$ means NO smearing effect and the smearing effect increases
as $\alpha$ gets small.
}
\label{efficiency}
\end{center}
\end{figure}

Table \ref{table2} is the list of radiation energy which is obtained by
adding up the bolometric luminosity from the time of collision
to around 300 days since the collision.
The model D3 is excluded because we do not have the entire numerical LC.
We also show the parameter $\alpha$
which is derived by using Equation (\ref{ef4}).
The efficiency $E_\mathrm{rad}/E_\mathrm{ej}$ to convert
SN kinetic energy to radiation is plotted in Figure \ref{efficiency}
as a function of
$M_\mathrm{CSM}/\left(M_\mathrm{ej}+M_\mathrm{CSM}\right)$
with the results obtained by \citet{vanmarle2010}.

At the high $M_\mathrm{CSM}/\left(M_\mathrm{ej}+M_\mathrm{CSM}\right)$
region, our standard $B_q=1$ results follow the line of $\alpha=0.5$.
This means that the efficiency to convert the kinetic energy
to radiation is reduced by 50\%.
On the other hand, the results of \citet{vanmarle2010} follow
the $\alpha=1$ line and the effect of multidimensional
instabilities is not significant.
Although \citet{vanmarle2010} use a three-dimensional code
and multidimensional instabilities are included in principle,
their approximated way to treat the radiation and
limited spatial resolution may have prevented multidimensional
instabilities from growing.

As $M_\mathrm{CSM}/\left(M_\mathrm{ej}+M_\mathrm{CSM}\right)$
becomes lower, the results start to deviate from the constant
$\alpha$ line. This is because $M_\mathrm{CSM}$ gets very small and
most of the ejecta is not affected by the interaction, i.e.,
the assumption $M_\mathrm{colej}=M_\mathrm{ej}$ is no longer valid
and Equation (\ref{ef4}) should not be used.
We should use Equation (\ref{ef3}) instead.
As $M_\mathrm{colej}\ll M_\mathrm{ej}$ in this regime,
the efficiencies tend to
be higher than the values obtained from Equation (\ref{ef4}).

The combinations of $E_\mathrm{ej}$ and $M_\mathrm{ej}$
which give a similar
$\left[M_\mathrm{sCSM}/\left(M_\mathrm{colej}+M_\mathrm{sCSM}\right)\right]
\frac{1}{2}M_\mathrm{colej}v_\mathrm{ej}^2$
are expected to result in similar LCs and they are degenerated.
Thus, it is difficult to constrain the exact value for $M_\mathrm{ej}$
and $E_\mathrm{ej}$ from LCs.
This is clearly seen in the models E1 and E2 in Figure \ref{RLC100}.
Both the models have similar LCs.
The CSM of the two models is exactly the same but $M_\mathrm{ej}$
and $E_\mathrm{ej}$ are different.
Although $M_\mathrm{colej}<M_\mathrm{ej}$ in the model E1
and $M_\mathrm{colej}\simeq M_\mathrm{ej}$ in the model E2,
$M_\mathrm{colej}$ and $\frac{1}{2}M_\mathrm{colej}v_\mathrm{ej}^2$
are happened to be
similar in the two models with the similar $M_\mathrm{sCSM}\simeq M_\mathrm{CSM}$.
Thus, the two models
have similar $\left[M_\mathrm{sCSM}/\left(M_\mathrm{colej}+M_\mathrm{sCSM}\right)\right]
\frac{1}{2}M_\mathrm{colej}v_\mathrm{ej}^2$
and result in the similar LCs.

\begin{table}
\centering
\begin{minipage}{70mm}
\caption{Conversion Efficiency from Kinetic Energy to Radiation Energy}
\label{table2}
\begin{tabular}{ccccc}
\hline
Name & $E_\mathrm{rad}$ & $E_\mathrm{rad}/E_\mathrm{ej}$ & 
$\frac{M_\mathrm{CSM}}{M_\mathrm{ej}+M_\mathrm{CSM}}$ & $\alpha$ \\
     &  $10^{51}$ erg   & & &   \\
\hline
A1 & 13   & 0.25  & 0.52 & 0.48 \\
A2 & 12   & 0.24  & 0.33 & 0.72 \\
B1 & 0.46 & 0.046 & 0.040 & 1.1 \\
B2 & 8.7 & 0.29   & 0.57 & 0.50 \\
C1 & 14  & 0.28   & 0.41 & 0.68 \\
D1 & 2.7 & 0.27   & 0.52 & 0.51 \\
D2 & 2.2 & 0.22   & 0.47 & 0.47 \\
E1 & 2.6 & 0.086  & 0.14 & 0.63 \\
E2 & 3.0 & 0.30   & 0.62 & 0.48 \\
F1 & 2.9 & 0.29   & 0.64 & 0.46 \\
D4 & 1.7 & 0.17   & 0.38 & 0.46 \\
D5 & 1.6 & 0.32   & 0.64 & 0.50 \\
D6 & 3.4 & 0.34   & 0.47 & 0.72 \\
D7 & 0.89 & 0.089 & 0.47 & 0.19 \\
\hline
\end{tabular}
\end{minipage}
\end{table}

\subsection{Origin of the Plateau Phase}\label{plateau}
There exists a plateau in the LC of SN 2006gy at around 200 days.
None of our models is succeeded in producing the plateau.
This is because the remaining CSM at these epochs is 
too thin to affect the LC.
Note that the LC observations at later epochs reject
the possibility to explain this plateau by \Ni\ heating (see also
Section \ref{sec:ni}).
There are several other possible ways
to explain the plateau.
One possibility is the recombination in the SN ejecta.
Because we use the simplified SN ejecta structures,
our results of LC calculations
after the photosphere gets inside of the SN ejecta
are beyond the applicability of our simple models.
Increasing the SN ejecta mass may also help
because a plateau phase can be longer with larger hydrogen mass,
although the conversion efficiency from kinetic energy to radiation
is also affected at the same time (Section \ref{sec:eff}).
By putting more realistic SN ejecta with realistic hydrogen-rich envelopes
or more massive SN ejecta,
the recombination wave may stay the envelope for a while
and may end up with the plateau phase, as is the case in
Type IIP SNe \citep[see, e.g.,][]{kasen2009}.
We note that the blackbody temperatures of these epochs are $\simeq 6000$ K and
they are consistent with this scenario.
Recombination may also occur in the shocked CSM or dense cool shell.

Light echoes from the remaining CSM may also play a role.
The LC after this plateau phase remains almost constant
for more than 200 days, although the luminosity is about 10 times
smaller \citep{miller2010}.
Thus, it is possible that there existed another CSM component which caused
the echoes at around 200 days and shocked away when SN 2006gy was
behind the Sun.

Smearing may also be relevant to the plateau.
In the models in Appendix \ref{appendix},
we can see that the plateau can appear or dissapear depending on
the degree of the smearing. Less smearing makes the cool shell denser
and photosphere can stay there for a longer time,
possibly making the plateau phase.

\begin{figure}
\begin{center}
 \includegraphics[width=\columnwidth]{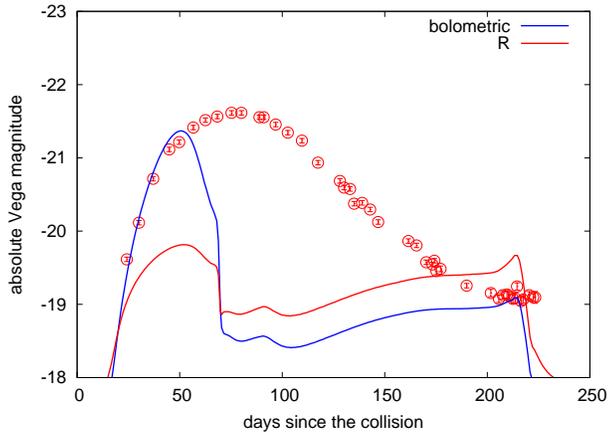}
  \caption{
Observed $R$ band LC and the bolometric and $R$ band LCs
calculated with the parameter shown in \citet{chatzopoulos2011}.
The observational points are shifted arbitrarily to match the
numerical result in the figure.
}
\label{fig:semi}
\end{center}
\end{figure}

\subsection{Progenitor of SN 2006gy}\label{progenitor}
As the properties of the progenitor system which can be obtained from
the LC modeling strongly depends on CSM, it is difficult to get information
on the progenitor from the LC of SN 2006gy.
The CSM properties and SN ejecta properties are degenerated.
However, we can get some indications for it.
As the origin of the luminosity is the kinetic energy of
SN ejecta, the kinetic energy should be converted
to radiation efficiently.
If $M_\mathrm{CSM}\ll M_\mathrm{ej}$,
the conversion efficiency
$\alpha M_\mathrm{CSM}/\left(M_\mathrm{ej}+M_\mathrm{CSM}\right)$
is so small that the kinetic energy cannot be converted efficiently enough
to explain the LC of SN 2006gy.
Thus, the mass of the CSM should be close to or larger than $M_\mathrm{ej}$.

According to our modeling,
$M_\mathrm{CSM}$ is required to be $\sim 10~M_\odot$
and this means that $M_\mathrm{ej}$ is expected to be
$\sim 10~M_\odot$ or less.
This indicates that the total mass of the system
well exceeds $10~M_\odot$ and the progenitor of SN 2006gy
should be a very massive star.
In addition, the progenitor should lose $M_\mathrm{CSM}$
within $\sim 10$ years before the explosion.
Our models for SN 2006gy have $M_\mathrm{CSM}\simeq 18~M_\odot$
and it may be difficult for red supergiants (RSGs) to have such mass
loss because of the following reason:
To have CSM with $18~M_\odot$,
the zero-age main-sequence mass of RSGs should be very large
but such massive stars suffer more from the radiation driven wind
during their main-sequence phase because
of their large luminosities. Thus, losing most of their mass
only just before their explosions might be difficult.
However, extensive mass loss of RSGs is suggested by 
many authors
\citep[e.g.,][]{vanloon2005,vanbeveren2007,smith2009,boyer2010,yoon2010,moriya2011b,georgy2011}
and
it is still possible that very massive RSGs lose
$\sim 10~M_\odot$ just before their explosions due to, e.g.,
pulsations \citep[e.g.,][]{yoon2010,heger1997,li1994},
dust \citep[e.g.,][]{vanloon2005}, or
g-mode oscillations (\citealt{quataert2012}, see also \citealt{arnett2011}).

Another possible progenitor of SN 2006gy is a very massive star in
the luminous blue variable (LBV, see, e.g., \citealt{humphreys1994})
phase which is suggested to be
the progenitor of SN 2006gy \citep[e.g.,][]{smith2007b}.
LBVs experience extensive mass loss and
some of them, e.g., $\eta$ Carinae, have $\sim 10~M_\odot$ CSM \citep[e.g.,][]{smith2006}.
Their typical wind velocities are also consistent with
the wind velocities estimated from the narrow P-Cygni H$\alpha$ profiles in SN
2006gy \citep[e.g.,][]{smith2010}.
Although LBVs are theoretically considered to be on the way to Wolf-Rayet stars
and do not explode \citep[e.g.,][]{crowther2007,vink2009},
the progenitor of Type IIn SN 2005gl is
found to be an LBV \citep{gal-yam2007,gal-yam2009b}.
Several other Type IIn SNe are also suggested to have
evidences for LBV progenitors \citep[e.g.,][]{kotak2006,trundle2008,smith2011,kiewe2012}.

A shell created due to the interaction between the RSG wind and the Wolf-Rayet wind
is another possible way to have a massive CSM \citep[e.g.,][]{dwarkadas2011}.
Alternatively, a shell created by pulsational instability
can be followed up by the SN ejecta, instead of the ejecta
of the next pulse as suggested by \citet{woosley2007}.
Some binary interaction may cause extensive mass loss \citep[e.g.,][]{hachisu2008}
but binary systems have not been
considered deeply as a possible progenitor of SLSNe yet
\citep[see][]{chevalier2012b,soker2012}.
The collision of massive stars in a dense
stellar cluster can make a massive star surrounded by a massive
CSM and it may also result in SN 2006gy-like SLSNe
(\citealt[][]{portegies2007}, see also \citealt{pan2011a}).

With the condition that $M_\mathrm{ej}$ is
similar to or less than $M_\mathrm{CSM}$,
the conversion efficiency of kinetic energy to radiation
(Equation (\ref{bare})) is expected to be $\simeq 50$ \% at most (Section \ref{sec:eff}).
As the radiation energy emitted by SN 2006gy exceeds $2\times10^{51}$~erg,
the SN ejecta should have more than $\simeq 4\times 10^{51}$~erg.
Thus, the SN explosion inside should be very energetic.
As the energy of our models is comparable to those of energetic broad-line Type Ic SNe
whose progenitors are suggested to be very massive
\citep[e.g.,][]{nomoto2011},
the estimated high energy may also indicate that the progenitor
mass is rather close to those of LBVs.
Note, however, that the host galaxy of SN 2006gy is
not metal-poor \citep[e.g.,][]{ofek2007}
while broad-line Type Ic SNe
appear more preferentially in low metallicity environments
\citep[e.g.,][]{arcavi2010,modjaz2011,sanders2012}.
In addition, the late time spectra of SN 2006gy are not similar to
those of broad-line Type Ic SNe \citep{kawabata2009},
although the late time spectra of Type Ic SLSN 2010gx show such features
\citep{pastorello2010}.

\begin{figure*}
\begin{center}
 \includegraphics[width=\columnwidth]{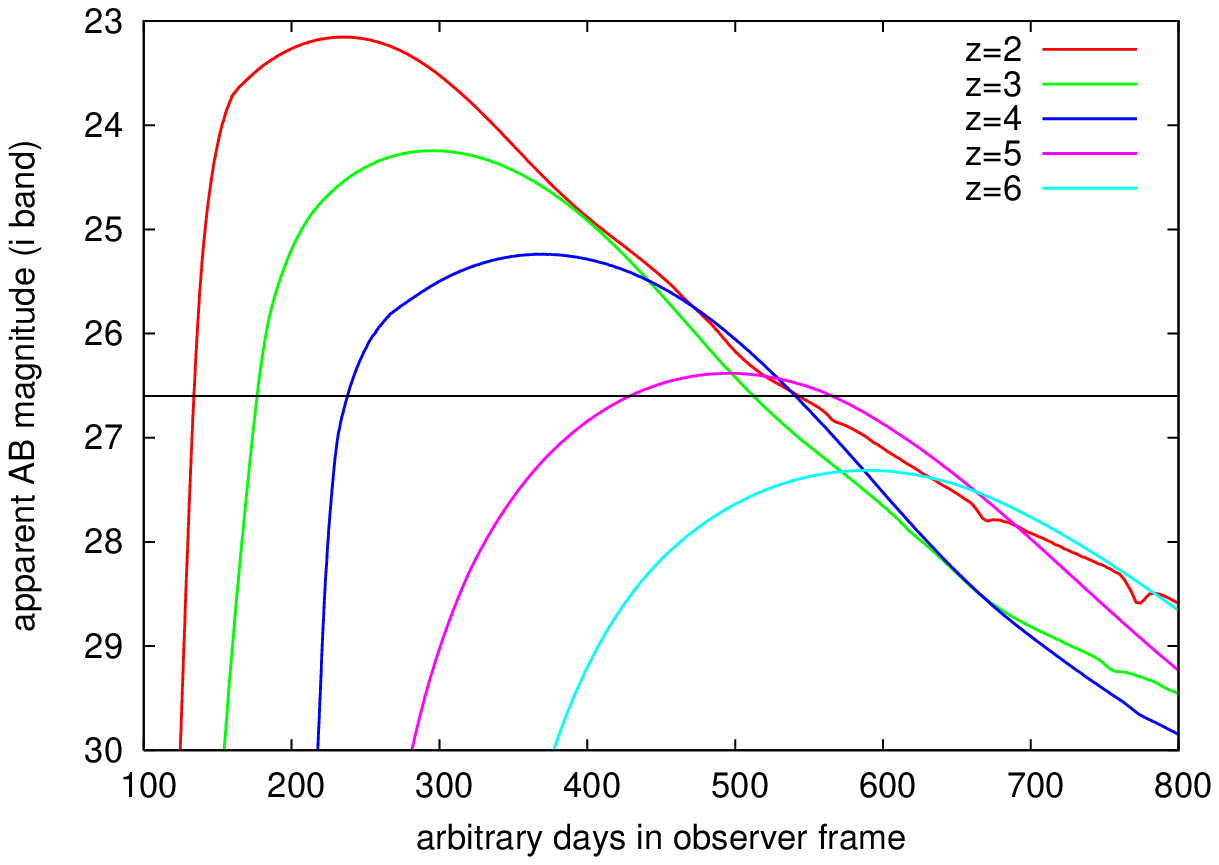}
 \includegraphics[width=\columnwidth]{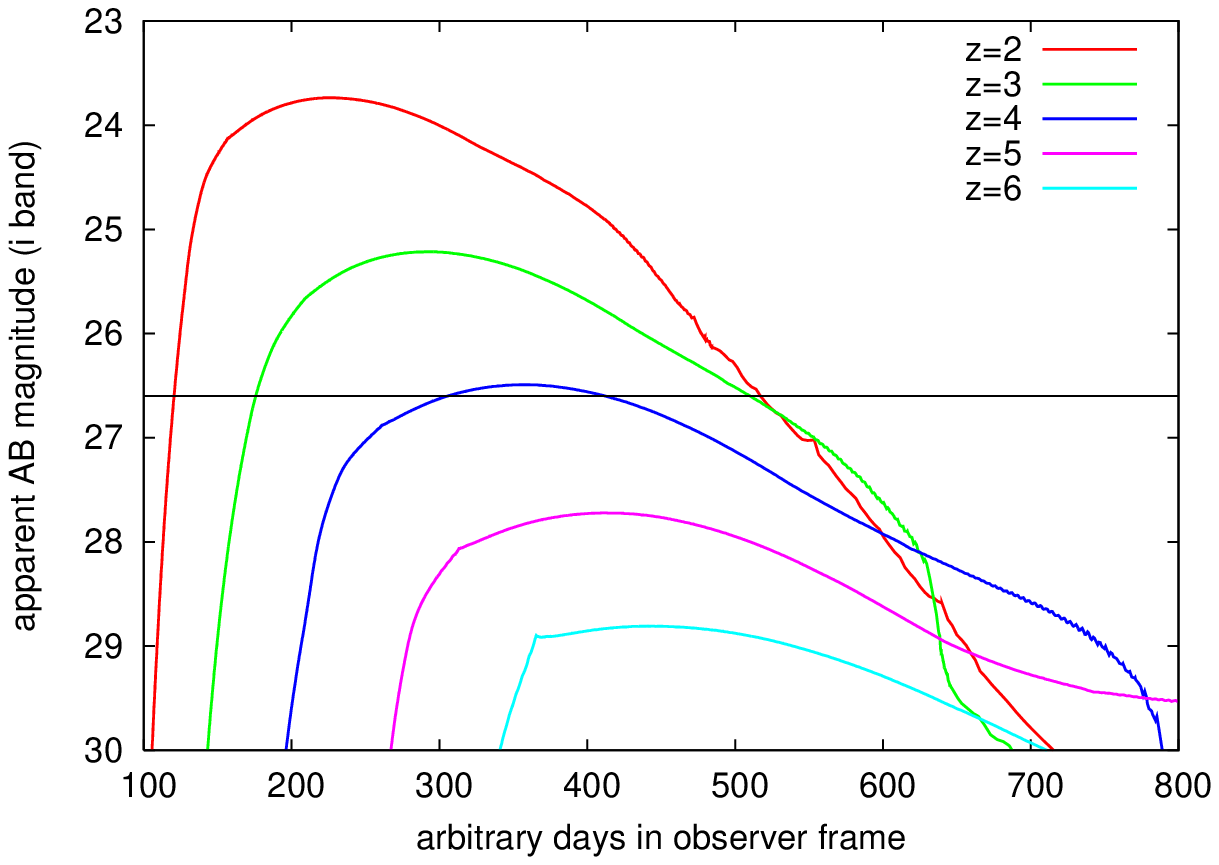}
  \caption{
$i$ band LCs of the models D2 ($w=5$, left) and F1 ($w=0$, right) at several redshifts.
The horizontal line is the planed $i$ band limiting magnitude of
Subaru/HSC Ultra-Deep survey at one epoch (26.6 mag).
}
\label{highzoptical}
\end{center}
\end{figure*}

\begin{figure*}
\begin{center}
 \includegraphics[width=\columnwidth]{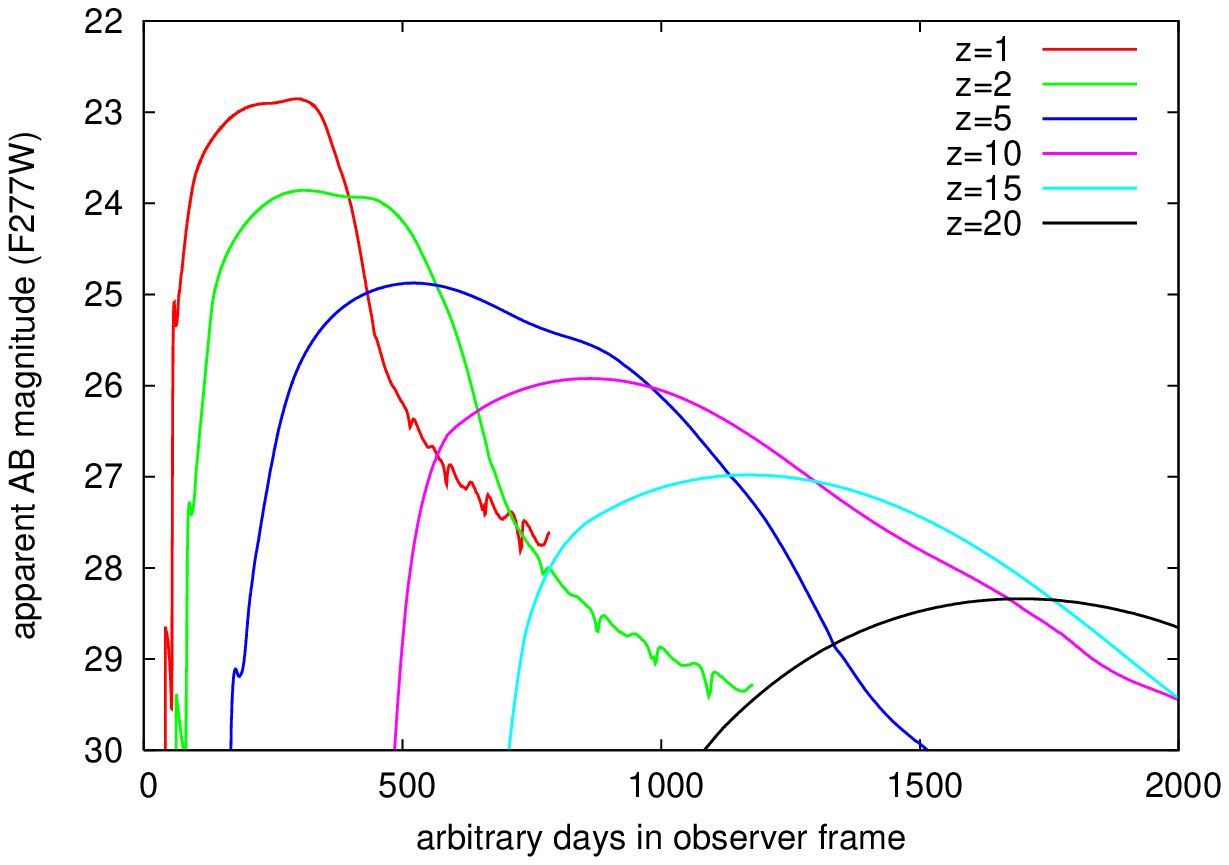}
 \includegraphics[width=\columnwidth]{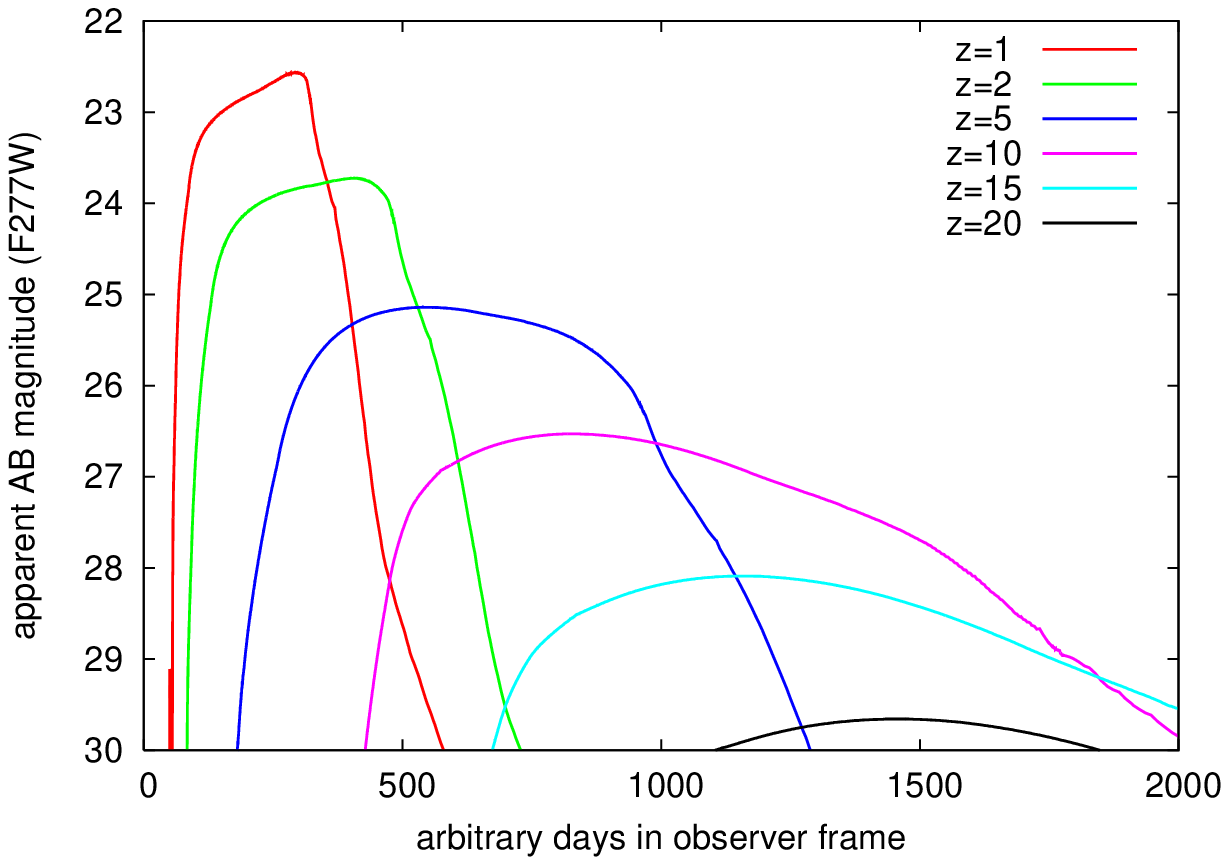}
  \caption{
F277W band LCs of the models D2 ($w=5$, left) and F1 ($w=0$, right) at several redshifts.
}
\label{highznir}
\end{center}
\end{figure*}

\subsection{Comparison to Semi-Analytic Model}\label{semi}
Recently, \citet{chatzopoulos2011} proposed a semi-analytic model
of LCs powered by the shock interaction.
The model involves several simplifications but the overall features
predicted by the model are shown to match some numerical results.
They show a LC model of SN 2006gy and we
perform LC calculations with the same parameters which
are obtained by them.
The parameters are $\delta=0$, $n=12$, $w=0$,
$E_\mathrm{ej}=4.4\times10^{51}$~erg,
$M_\mathrm{ej}=40~M_\odot$,
$R_i=5\times 10^{14}$~cm, and
$R_o=2.5\times 10^{15}$~cm
(corresponds to $M_\mathrm{CSM}=5~M_\odot$ with
a constant CSM density $1.5\times 10^{-13}$~$\mathrm{g~cm^{-3}}$).
Following the result of \citet{chatzopoulos2011},
we put $2~M_\odot$ of \Ni\ at the center of the ejecta 
but the value is too small to affect the main part of the LC (Section \ref{sec:ni}).
Except for the central region, the composition is set as the solar metallicity.
We use our standard $B_q=1$.

Figure \ref{fig:semi} shows the result of the numerical calculation.
Overall, the parameters suggested by \citet{chatzopoulos2011}
do not result in a similar LC to that of SN 2006gy.
The peak luminosity of the bolometric LC is close to that of SN 2006gy
but the duration is much shorter that that of SN 2006gy.
In addition, the $R$ band peak luminosity is much smaller than
that of SN 2006gy because the photospheric temperature is
much higher than that of SN 2006gy around the peak and the model
has much bluer spectra.
The duration can be longer if we use a smaller $B_q$
but our results shown in Figure \ref{BQ}
imply that it is difficult to make the duration
two times longer than the $B_q=1$ LC to match the LC of SN 2006gy
by just making the $B_q$ small.
What is more, changing $B_q$ does not improve the color of
the LC and the $R$ band LC is expected to remain much
fainter than the observed $R$ band LC.
The plateau phase after the drop in the LC is due to the
recombination in $40~M_\odot$ SN ejecta inside.
The duration of the plateau phase should actually
be much shorter than that in Figure \ref{fig:semi},
as the entire SN ejecta is composed of the solar metallicity in the model
(except for the central \Ni) and we use an approximate density structure
for the SN ejecta.

There are several possible reasons for the discrepancy.
The semi-analytic model
assumes that the thermal energy gained by the forward shock is
always released at the center of the CSM because
the assumption is required to treat the transport equations
analytically.
However, this assumption keeps the diffusion time
of the photons from the forward shock constant
and the diffusion time is fixed with the initial value.
In reality, the forward shock travels outward
and the diffusion time decreases with time
as the remaining unshocked wind decreases.
This effect leads to the overestimation of
the duration of the LCs in the semi-analytic model.
This is presumed to be the
main reason why the duration of the LC obtained
by the numerical calculation is much shorter than that obtained by
the analytical model.
This indicates that the semi-analytic model 
should not be applied to the system with CSM of
a large initial diffusion time in which the kinetic energy
of shock waves is the main source of radiation.

Another possible reason is that the energy released by the 
reverse shock is overestimated in the semi-analytic model.
In the semi-analytic model, the self-similar solution of
\citet{chevalier1982,nadezhin1985} is used as the evolution of the hydrodynamical structure.
However, in reality, the effect of cooling which is not taken into
account in the adiabatic self-similar solution is so strong
in the case of SLSNe powered by the shock interaction that
a thin cool dense shell is created between the SN ejecta and the dense
CSM.
Thus, the reverse shock could not travel as fast as expected
from the adiabatic self-similar solution and rather stays close to
the forward shock.

In summary, many important effects which are essential in modeling
the LC powered by the shock interaction between SN ejecta and a
dense CSM with a large photon diffusion time
lack in the semi-analytic model 
and it may not be appropriate to use it for the modeling of SLSNe
powered by such a strong interaction.

\subsection{High-Redshift Type IIn Superluminous Supernovae}
SLSNe powered by the shock interaction are
not only brighter but also bluer than other SNe.
Thus, SLSNe are expected to be a good probe to study the high-redshift
Universe. Especially, as their progenitors are
expected to be very massive (Section \ref{progenitor}),
SLSNe can provide us with the information of very massive stars
in the early Universe.
Some high-redshift Type IIn SNe are already detected 
up to $z=2.36$ (\citealt{cooke2009}, see also \citealt{barton2010}).

With upcoming optical deep surveys with, e.g.,
Subaru/Hyper Suprime-Cam (HSC) \citep{miyazaki2006},
SLSNe powered by the interaction can be detected
up to $z\sim5$ \cite[e.g.,][]{tanaka2012,cooke2008} where as
PISNe can be detected up to $z\sim2$
\citep[e.g.,][]{pan2011a}.
Upcoming NIR surveys by, e.g.,
James Webb Space Telescope (JWST)\footnote{\url{http://www.jwst.nasa.gov/}},
Euclid\footnote{\url{http://sci.esa.int/euclid}},
or
Wide-field Imaging Surveyor for High-redshift (WISH)\footnote{
\url{http://www.wishmission.org/en/index.html}},
can approach to much higher
redshifts \citep[e.g.,][]{tanaka2012,pan2011b,hummel2011,scannapieco2005}.
Here, we show how the LC of SN 2006gy is observed if it appears
at high redshifts.
This can be used for the identification of candidate transients
obtained in upcoming optical and NIR surveys.

Figure \ref{highzoptical} is the $i$ band LCs of the model
D2 $(w=5)$ and F1 $(w=0)$ for several redshifts.
As the model F1 has similar color to SN 2006gy
at around the LC peak, the F1 model provides better estimate
for the high-redshift SLSN LCs.
Figure \ref{highznir} is the LCs with the F277W filter\footnote{
\url{http://www.stsci.edu/jwst/instruments/nircam/instrumentdesign/filters/}}
which is one of the wide filters planned for JWST.
The central wevelength of the F277W filter is at 2.77 $\mu$m.
JWST can reach $\simeq 30$ mag and is able to detect SLSNe beyond $z=10$.
See \citet{tanaka2012} for the details of observational strategies
and estimated number of detections of SLSNe
with upcoming optical and NIR surveys.

\section{Summary and Conclusions}\label{conclusions}
We have shown that the interaction between SN ejecta and dense CSM
is a viable mechanism to power SLSNe such as SN 2006gy.
The interaction in the dense CSM accounts for the huge luminosity 
and the long duration of the SN 2006gy LC.
Shock breakout within the dense CSM is a key for the understanding of the
interaction-powered SLSNe.
Our canonical models have
$M_\mathrm{ej}=20~M_\odot$, $E_\mathrm{ej}=10^{52}$~erg, and
$M_\mathrm{CSM}=18~M_\odot$ $(w=5)$ or $15~M_\odot$ $(w=0)$
where the CSM is assumed to have a density profile of $\rho\propto r^{-w}$.
The corresponding average mass-loss rate of the progenitor
is about $0.4~M_\odot~\mathrm{yr^{-1}}$ if we assume that
the dense CSM originates from a $100~\mathrm{km~s^{-1}}$ wind.
Our steady mass-loss models $(w=2)$ fail to explain the SN 2006gy LC.
No \Ni\ is required to explain the early LC of SN 2006gy.

It is difficult to break the degeneracy among
$M_\mathrm{ej}$, $E_\mathrm{ej}$, and $M_\mathrm{CSM}$.
One can obtain constraints on the progenitor of SN 2006gy
based on the efficiency,
as the conversion efficiency of the SN kinetic energy to radiation
becomes high when
$M_\mathrm{ej}$ is comparable to or less than $M_\mathrm{CSM}$.
The progenitor of SN 2006gy should be a very massive star
because $M_\mathrm{CSM}= 18~M_\odot$ or $15~M_\odot$
As the conversion efficiency is $\simeq 50$ \% at most and 
the radiation energy emitted by SN 2006gy is more than $2\times 10^{51}$
erg, $E_\mathrm{ej}$ should be larger than $4\times 10^{51}$ erg.

We have also examined the effect of multidimensional
instabilities in the dense cool shell on the model LCs.
Such instabilities are expected to reduce the amount of kinetic energy
converted to radiation. 
Our LC modeling is based on a one-dimensional radiation
hydrodynamics code in which the multidimensional
instabilities are implemented only in an approximate way.
We have thus explore the effect qualitatively. 
Further studies on the multidimensinal effect(s),
perhaps using three-dimensional radiation hydrodynamics simulations,
are needed for better understanding of SNe powered by the interaction.

Finally, we have provided predictions from our model for 
high-redshift SLSNe. 
We have calculated the optical and NIR LC evolution in the observer frame.
The results can be used for the identification of SLSNe
appeared at high redshifts in the future transient surveys.

The existence of the very massive CSM close to the progenitor which
is required to explain the LC of SN 2006gy challenges the current
understanding of the stellar mass loss.
The better understanding of SNe powered by the interaction
will lead us to the better understanding of the mass loss
mechanisms of massive stars.
This is a critical key to reveal the evolution and the fates of massive stars
which are a fundamental component in the Universe.

\section*{Acknowledgments} 
We thank the anonymous referee for the comments.
T.J.M. is supported by the Japan Society for the Promotion of Science
Research Fellowship for Young Scientists $(23\cdot5929)$.
S.I.B. is supported partly by the grants
of the Government of the Russian Federation (No 11.G34.31.0047),
RFBR  10-02-00249, 10-02-01398,
by RF Sci.~Schools 3458.2010.2  and  3899.2010.2,
and by a grant IZ73Z0-128180/1 of the Swiss National Science
Foundation (SCOPES).
All the numerical calculations were carried out on the general-purpose
PC farm at Center for Computational Astrophysics, CfCA, of National
Astronomical Observatory of Japan.
N.Y. acknowledges support by the Grants-in-Aid for Young Scientists
(S: 20674003) by the Japan Society for the Promotion of Science.
This research is supported in part
by a grant from the Hayakawa Satio Fund awarded by the Astronomical
Society of Japan. 
This research is also supported by World Premier International Research Center Initiative, MEXT, Japan.

\appendix

\section{Effect of the Smearing Term on Light Curves of
Pulsational Pair-Instability Supernovae}\label{appendix}
We show the results of our additional investigations
on the dependence of the smearing term $B_q$
on the results of LC calculations.
We show the results obtained from
the pulsational-pair instability SN models presented
by \citet{woosley2007} and show that the effect of $B_q$
on LCs is similar to our LC models.
This means that the effect of $B_q$ appeared in our models
is not unique to our models.

\begin{figure}
\begin{center}
 \includegraphics[width=\columnwidth]{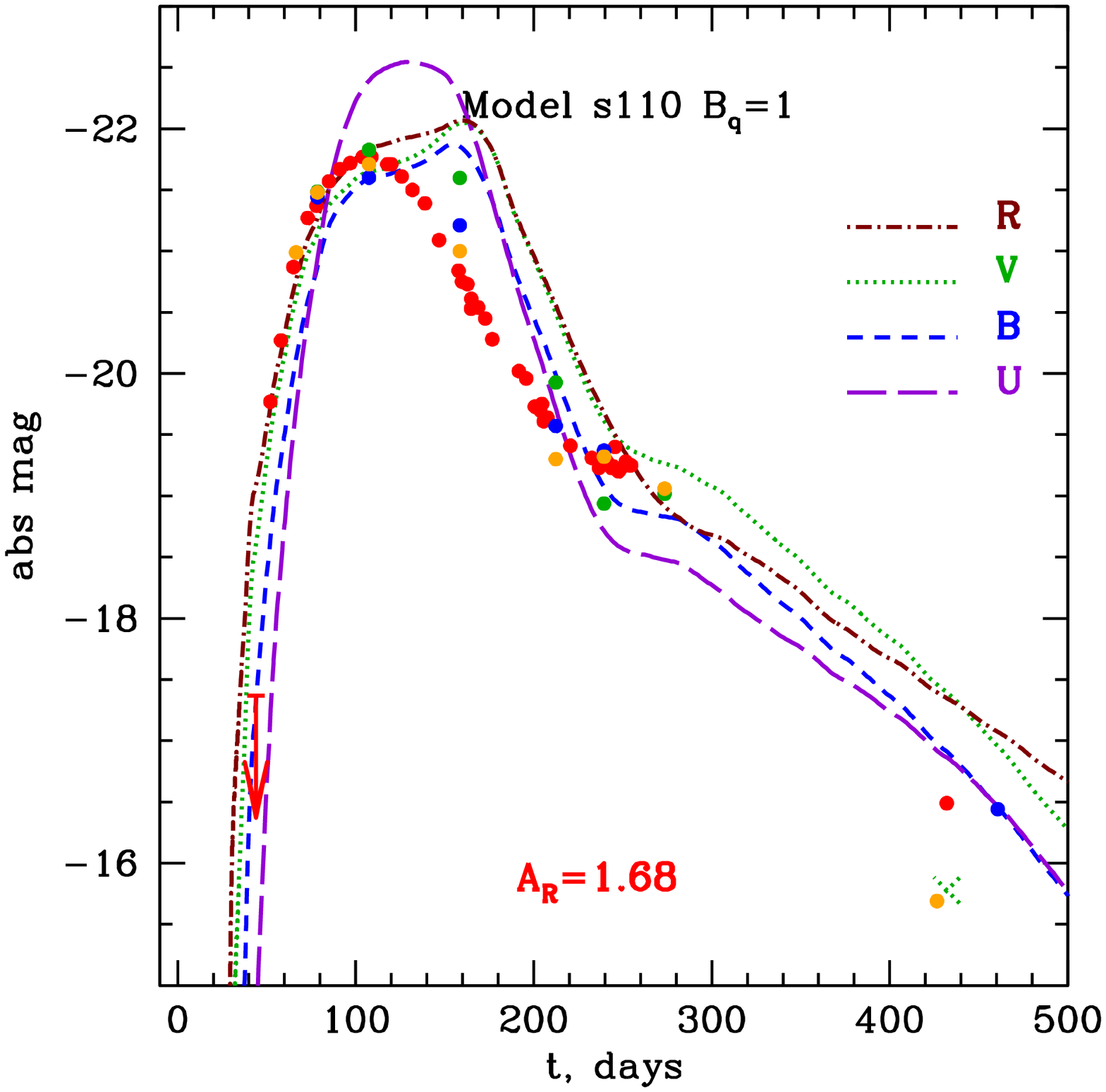}
  \caption{
LCs of a pulsational-pair instability model with $B_q=1$.
The kinetic energy of the second pulse is increased nine times 
as large as the original model.
Note that slightly higher absorption ($A_R=1.68$ mag)
is applied in the observational data points
than the value adopted in the other part of this paper ($A_R=1.3$ mag).
}
\label{w1}
\end{center}
\end{figure}
\begin{figure}
\begin{center}
 \includegraphics[width=\columnwidth]{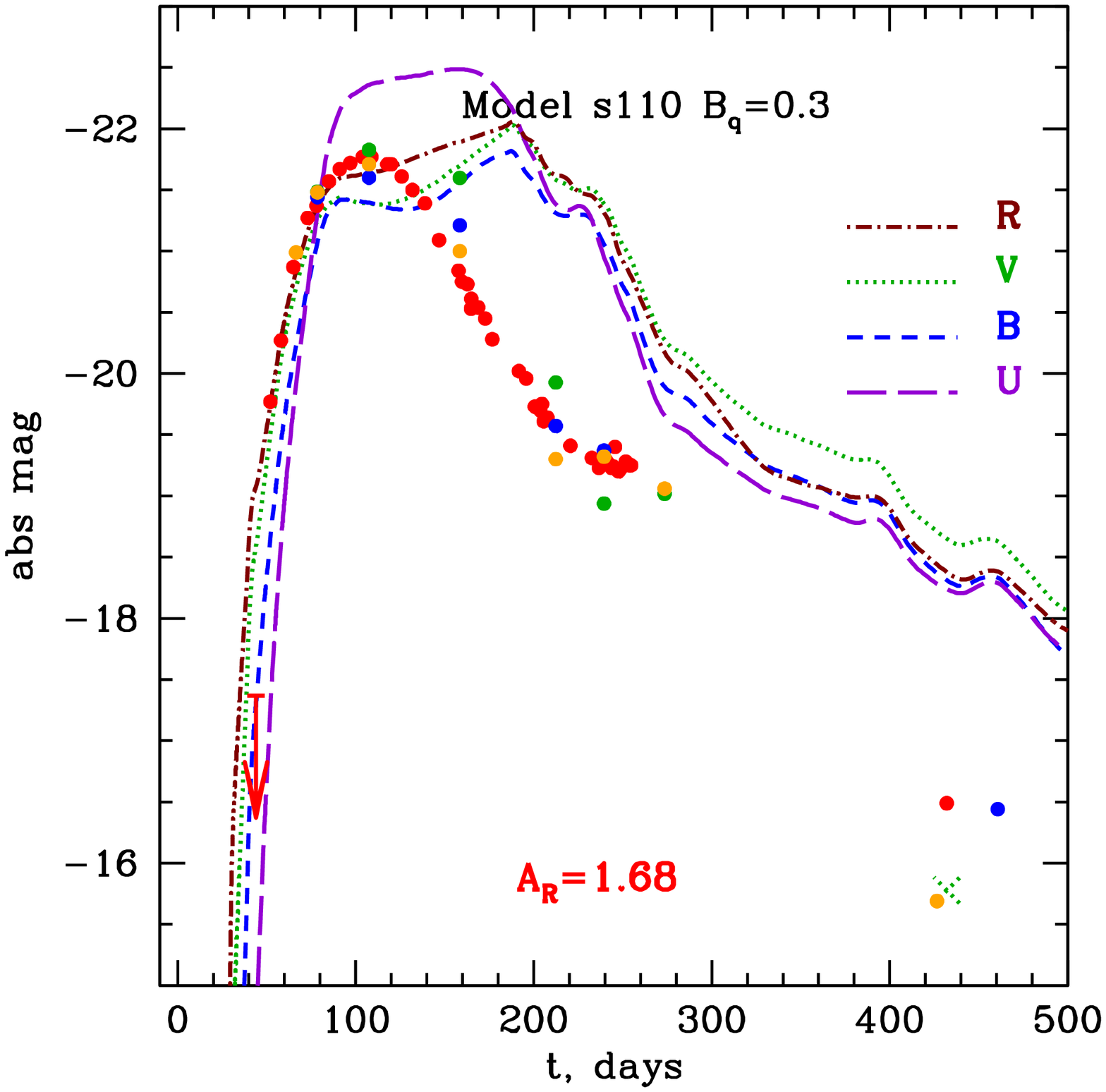}
  \caption{
The same as Figure \ref{w1} but $B_q=0.3$.
}
\label{w2}
\end{center}
\end{figure}
\begin{figure}
\begin{center}
 \includegraphics[width=\columnwidth]{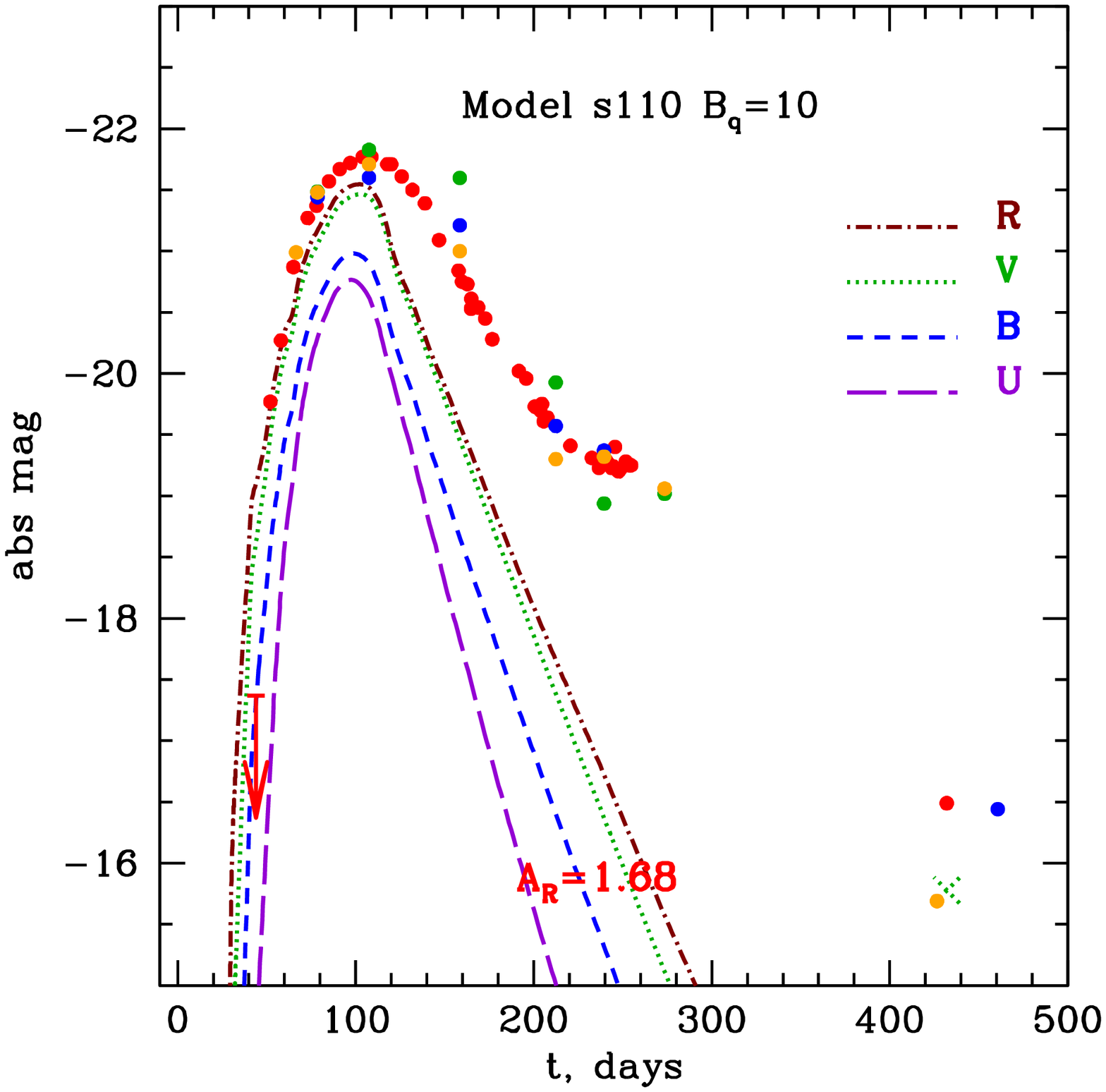}
  \caption{
The same as Figure \ref{w1} but $B_q=10$.
}
\label{w3}
\end{center}
\end{figure}

The models presented here are based on the same progenitor model
($M_\mathrm{ZAMS}=110~M_\odot$) shown in \citet{woosley2007}.
The shell ejected by the first pulsation ($24.5~M_\odot$)
is caught up by the materials ejected by the next pulsation ($5.1~M_\odot$).
\citet{woosley2007} have artificially increased
the kinetic energy of the materials ejected by the second pulsation
to obtain the better fit to the LC of SN 2006gy and 
their best LC model has four times as much kinetic energy as that of the original model.
The models shown here have more kinetic energy.
The kinetic energy of the second pulse is increased nine times
and the kinetic energy of the second pulse is $6.5\times 10^{51}$ erg.
The rest of the model is the same as those discussed in
\citet{woosley2007}.

Figures \ref{w1}, \ref{w2}, and \ref{w3} show the results
of the LC calculations with different $B_q$.
Comparing the results to those shown in Figure \ref{BQ}, 
we can see that the dependence of the results on $B_q$ is
basically the same.
This is because the source of the luminosity in
both the models is kinetic energy of the material
coming from inside and $B_q$ directly affects the conversion
efficiency of the kinetic energy to the radiation energy.
For the accurate modeling of the LCs resulting from 
the conversion of the kinetic energy to radiation,
we need more investigations on the multidimensional effects
during the interaction.

\label{lastpage}

\end{document}